# Magnetoresistance Effects in $SrFeO_{3-\delta}$: Dependence on Phase Composition and Relation to Magnetic and Charge Order

P. Adler[1,2], A. Lebon[1,3], V. Damljanović[1], C. Ulrich[1], C. Bernhard[1], A. V. Boris[1], A. Maljuk[1], C. T. Lin[1], and B. Keimer[1]

[1]*Max- Planck-Institut für Festkörperforschung, Heisenbergstrasse 1, D-70569 Stuttgart, Germany*

[2]*Institut für Anorganísche Chemie, Universität Karlsruhe (TH), Engesserstrasse, Geb. 30.45, D-76128 Karlsruhe, Germany*

[3]*Laboratoire de Physique des collisions électroniques et atomiques, 6 avenue Le Gorgeu 29285 Brest Cedex, France.*

Single crystals of iron(IV) rich oxides $SrFeO_{3-\delta}$ with controlled oxygen content ($0 \leq \delta \leq 0.31$) have been studied by Mössbauer spectroscopy, magnetometry, magnetotransport measurements, Raman spectroscopy, and infrared ellipsometry in order to relate the large magnetoresistance (MR) effects in this system to phase composition, magnetic and charge order. It is shown that three different types of MR effects occur. In cubic $SrFeO_3$ ($\delta = 0$) a large negative MR of 25% at 9 T is associated with a hitherto unknown 60 K magnetic transition and a subsequent drop in resistivity. The 60 K transition appears in addition to the onset of helical ordering at ~130 K. In crystals with vacancy-ordered tetragonal $SrFeO_{3-\delta}$ as majority phase ($\delta$ ~0.15) a coincident charge/antiferromagnetic ordering transition near 70 K gives rise to a negative giant MR effect of 90% at 9 T. A positive MR effect is observed in tetragonal and orthorhombic materials with increased oxygen deficiency ($\delta$ = 0.19, 0.23) which are insulating at low temperatures. Phase mixtures can result in a complex superposition of these different MR phenomena. The MR effects in $SrFeO_{3-\delta}$ differ from those in manganites as no ferromagnetic states are involved.

## I. INTRODUCTION

The complex electronic phase diagrams of many transition metal (TM) oxide systems, for instance the intensely studied perovskite-type manganites $Ln_{1-x}A_xMnO_3$ [1,2], reflect a situation where the width of the conduction band, the electron correlation energy, and the strength of the electron-phonon coupling are of comparable magnitude. If two electronic phases are separated by a first-order transition, phase coexistence is often observed over a range of compositions, and the volume fractions can be changed by varying the composition, temperature, pressure, or magnetic field. This situation is encountered in the colossal magnetoresistance (CMR) manganites, where the ferromagnetic metallic state is stabilized by a magnetic field with respect to competing charge ordered antiferromagnetic or paramagnetic insulating states [3,4]. As a baseline for a description of the properties of the CMR manganites, it is helpful to consider the electronic structure of the prototype undoped parent compound $LaMnO_3$, which contains only $Mn^{3+}$ ions with a $t_{2g}^3e_g^1$ high spin configuration. Since the lowest energy optical excitations in $LaMnO_3$ were assigned to intersite d-d transitions [5], $LaMnO_3$ can be considered as a Mott-Hubbard type insulator. Both, $t_{2g}$ and $e_g$ electrons are localized, and the $t_{2g}^3e_g^1$ electron configuration gives rise to a cooperative Jahn-Teller effect. This is the origin of orbital ordering ($T_{OO}$ ~ 750 K) [6] and of the type-A antiferromagnetic spin structure ($T_N$ = 140 K) [7]. Hole doping of $LaMnO_3$ leads to the formation of $Mn^{4+}$ ($t_{2g}^3$) ions. The double exchange (DE) mechanism, which favors delocalization of the $e_g$ electrons if the $t_{2g}$ spins are ferromagnetically aligned, then stabilizes a ferromagnetic metallic state. Depending on the composition and structural details, however, various charge-ordered antiferromagnetic states have also been observed.

In order to improve our understanding of the physical properties of the manganites, it is interesting to consider iron(IV) based oxides, because the $Fe^{4+}$ ion is isoelectronic with the $Mn^{3+}$ ion. In contrast to $LaMnO_3$ the prototype Fe(IV) oxide $SrFeO_3$ is metallic and exhibits a cubic perovskite structure at all temperatures, without any evidence of a Jahn-Teller effect [8,9]. The different properties of Fe(IV) compared to Mn(III) oxides reflect the strongly enhanced covalency of iron(IV)-oxygen bonding, which is in agreement with the chemical trend that the M 3d orbitals are stabilized from the left to the right in a TM series and with increasing oxidation state for a given M. This is corroborated by band structure calculations of $SrFeO_3$ [10] as well as by comparative band structure calculations of $CaFeO_3$ and $LaMnO_3$ [11]. The larger covalency leads to a larger width of the conduction band which is formed from the $\sigma^*$ (Fe 3d – O2p) antibonding $e_g$ orbitals. The increased itinerancy of the $e_g$ electrons explains the absence of the Jahn-Teller effect in $SrFeO_3$ and related compounds. Basic electronic structure parameters of $SrFeO_3$ were derived from the analysis of Fe 2p core level spectra within a cluster configuration interaction model [12]. Within this approach, the strong covalency of the Fe-O bonding results in a negative effective charge transfer energy Δ, which is the energy difference between the lowest lying states arising from the $d^4$ and $d^5L$ multiplets. This means that the electronic ground state of Fe(IV) oxides is dominated by the $d^5L$ rather than the ionic $d^4$ configurations. Here L denotes a hole in the oxygen 2p orbitals.

Overall, the electronic situation in single-valent $SrFeO_3$ appears to be comparable to that in mixed-valent manganites. In both cases the DE mechanism favors the delocalization of the $e_g$ electrons, which are coupled to the localized $t_{2g}^3$ subshell. In contrast to the manganites, no doping is required, and the itinerancy of the $e_g$ electrons is bandwidth-driven in $SrFeO_3$. However, the charge carriers have a larger O 2p hole character. While it might be expected that the DE mechanism results in a ferromagnetic metallic state for $SrFeO_3$, the compound actually undergoes a phase transition from a paramagnetic to a helicoidal magnetic state near 130 K [13,14]. This was originally explained by a competition between antiferromagnetic superexchange and DE interactions. Recently this view has been challenged, and it was shown by model calculations that for negative Δ the DE mechanism alone is capable of stabilizing an incommensurate helical spin structure [15]. The metallic state found in $SrFeO_3$ becomes unstable with respect to a charge disproportionation of $Fe^{4+}$ into $Fe^{(4-\delta)+}$ ("$Fe^{3+}$") and $Fe^{(4+\delta)+}$ ("$Fe^{5+}$"), if the $\sigma*$ bandwidth is somewhat reduced. This situation occurs in the distorted perovskite $CaFeO_3$ [16], where an intricate interplay between lattice distortions and correlation effects [17] leads to charge ordering ($T_{CO}$ = 290 K) [18] and the opening of a band gap. Remarkably, the helical spin structure ($T_N$ = 115 K) is not much influenced by the insulator-metal transition driven by replacing Sr with Ca. This can also be understood as a consequence of the negative Δ in these materials [15].

The above considerations show that in iron(IV) based oxides spin-dependent electron delocalization processes as well as electron-lattice interactions are important. This has motivated a search for CMR effects akin to those observed in the manganites. It turned out that a successful route to large MR effects in Fe(IV) materials is the partial substitution of iron by cobalt or nickel, which stabilizes ferromagnetic ordering. MR effects up to -65% at 4 K and 5 T have been achieved in the systems $SrFe_{1-x}Co_xO_{3-\delta}$ [19,20], $Sr_3Fe_{2-x}(Co,Ni)_xO_{7-\delta}$ [21 - 25], and $Sr_{2/3}La_{1/3}Fe_{1-x}Co_xO_{3-\delta}$ [26]. Recently, Zhao et al. reported a sizable MR effect of up to -15% at 4 K and 9 T even for unsubstituted polycrystalline $SrFeO_{2.95}$ [27]. This effect is apparently related to an anomaly in the resistivity data near 55 K, which can be shifted by a magnetic field. The resistivity anomaly in turn appeared to be correlated with an anomaly near 65 K in magnetic susceptibility data and the appearance of paramagnetic $Fe^{(3+\delta)+}$ sites in the Mössbauer spectra for T $\geq$ 80 K (which is below the Néel temperature in the $SrFeO_{2.95}$ sample). In the light of earlier neutron diffraction experiments on $SrFeO_{2.90}$ [14], it was proposed that a modification of the pitch angle of the helical spin structure or a magnetic-field induced change from the helical to a conical spin structure could be reasons for the observed magnetotransport effects.

In contrast to manganites and the cobalt- or nickel-substituted ferrates, the MR effect in $SrFeO_{2.95}$ does not involve a ferromagnetic state. It is therefore of fundamental interest to understand its microscopic origin. Since previous experience with the manganites has shown that the coexistence of different phases can have a key influence on the physical properties [3,4], it is essential to consider the exact phase compositions in the investigated samples. It is known that the structural phase diagram of the $SrFeO_{3-\delta}$ system encompasses four phases, namely the stoichiometric cubic phase (C, $\delta$ = 0) as well as the oxygen vacancy ordered tetragonal (T, $\delta$ = 0.125, $Sr_8Fe_8O_{23}$), orthorhombic (O, $\delta$ = 0.25, $Sr_4Fe_4O_{11}$), and brownmillerite-type (B, $\delta$ = 0.5, $Sr_2Fe_2O_5$) phases separated by miscibility gaps [28]. Other compositions correspond to phase mixtures. In order to investigate the magnetotransport properties in the system $SrFeO_{3-\delta}$ in more detail and to establish the relation to magnetic and charge ordering, we have conducted a systematic study of a series of $SrFeO_{3-\delta}$ single crystals with different oxygen contents ($\delta \leq 0.25$). Some highlights of our work have already been communicated [29]. Here we report a comprehensive set of magnetotransport data, as well as Mössbauer, Raman, and optical reflectivity spectra, which enable us to give a detailed account of the relationship between the phase compositions, the magnetic and charge ordering transitions, and the magnetotransport properties. In particular, it is shown that depending on the phases present different types of magnetoresistance phenomena occur in this system. A large negative MR is observed in cubic $SrFeO_3$, which is associated with a transition seen near 55 K in the magnetic as well as in the resistivity data. A giant negative MR related to a combined magnetic/charge ordering transition near 70 K is found for $SrFeO_{2.85}$, a material with the tetragonal phase as the major component. Finally, a large positive MR effect at low temperatures is evident in samples with further increased oxygen deficiency. In the light of our results, we reconsider the data reported by Zhao et al. and argue that the properties of their sample should be interpreted in terms of a mixture of cubic and tetragonal phases.

The paper is structured as follows. In order to support the interpretation of our experimental findings, the most important structural features of the system $SrFeO_{3-\delta}$ are summarized in Section II. In Section III the experimental details are given. In Section IV Mössbauer, magnetotransport, and optical data are presented. Finally, Section V offers some Concluding Remarks.

## II. SUMMARY OF STRUCTURAL DATA

The known phases in the system $SrFeO_{3-\delta}$ can be described by the general formula $Sr_nFe_nO_{3n-1}$ with n = ∞, 8, 4, 2. Their crystal structures were studied previously at room temperature by powder neutron diffraction [28]. The n = ∞ member $SrFeO_3$ adopts the cubic perovskite structure (space group Pm-3m) with a three-dimensional network of corner-sharing regular $FeO_6$ octahedra. The other members of the series correspond to oxygen vacancy-ordered defect perovskites. The iron-oxygen distances are summarized in Table 1, the crystal structures of tetragonal $Sr_8Fe_8O_{23}$ (n = 8, space group I4/mmm) and orthorhombic $Sr_4Fe_4O_{11}$ (n = 4, space group Cmmm) are illustrated in Fig. 1. The iron-oxygen network in $Sr_8Fe_8O_{23}$ involves three different types of iron sites Fe1, Fe2, and Fe3 with a relative abundance of 1:2:1. The Fe1 sites are in an approximate square pyramidal environment of oxygen atoms, whereas the Fe2 and Fe3 sites are six-fold coordinated. Similar to the cubic perovskite structure, all polyhedra are connected via corner-sharing of oxygen atoms. However, oxygen deficiency leads to the formation of $Fe1O_5$ square pyramids, two of which form dimers via sharing of the O1 atom. The pyramids are connected only to distorted $Fe2O_6$ octahedra, which themselves are connected to all other types of polyhedra. The $Fe3O_6$ octahedra are nearly regular and form rows along the c-direction. Each of these rows is surrounded by four columns of $Fe2O_6$ units. The Fe3-O distances are very similar to those in $SrFeO_3$, and the Fe3 sites are readily attributed to $Fe^{4+}$. The Fe-O distances in the $FeO_5$ units of $Sr_8Fe_8O_{23}$ are more than 0.1 Å shorter than those in $Sr_3Fe_2O_6$ (4 × 1.98 Å, 1 × 1.92 Å) [30], the crystal structure of which contains only five-fold coordinated $Fe^{3+}$ ions. Thus the Fe1 sites are also assigned to $Fe^{4+}$, whereas the remaining Fe2 sites are attributed to the $Fe^{3.5+}$ species. The presence of the latter is supported by the intermediate Fe2-O distances and the Mössbauer spectra of $Sr_8Fe_8O_{23}$ [31,32].

Only two iron sites are present in the crystal structure of $Sr_4Fe_4O_{11}$ [28]. Again, dimers of $Fe1O_5$ units are formed. These are connected to rows of strongly distorted $Fe2O_6$ octahedra. The bond distances of the square pyramids are similar to those in $Sr_8Fe_8O_{23}$.

Therefore the Fe1 sites again correspond to $Fe^{4+}$, whereas the large Fe-O distances in the Fe2$O_6$ octahedra are in accord with an assignment as $Fe^{3+}$. This was concluded previously on the basis of bond valence calculations [28] whose validity has, however, been questioned in course of a recent neutron diffraction study on $Sr_4Fe_4O_{11}$ [33]. Considering a possible Jahn-Teller effect of $Fe^{4+}$ it was proposed that the distorted Fe2$O_6$ octahedra should be associated with the $Fe^{4+}$ sites and the square pyramids with the $Fe^{3+}$ sites. We discard this possibility as it is in clear contradiction to the known bond distances in perovskite-type $Fe^{3+}$ and $Fe^{4+}$ oxides (see above). Furthermore, due to strong covalency of iron-oxygen bonding the Jahn-Teller effect of $Fe^{4+}$ is usually suppressed.

Finally we mention the brownmillerite-type crystal structure of the pure $Fe^{3+}$ phase $Sr_2Fe_2O_5$ (n = 2 space group Icmm [28] or Ibm2 [34]) where alternating layers of tetrahedrally and octahedrally coordinated iron sites occur.

## III. EXPERIMENTAL DETAILS

The experiments were performed on high quality $SrFeO_{3-\delta}$ single crystals synthesized in a four-mirror type floating zone furnace [35]. The oxygen contents were determined to an accuracy of ~0.02 by thermogravimetry (TG) in a reducing Ar/$H_2$ atmosphere ($H_2$ content 5 at.%). As-grown oxygen-deficient crystals with $\delta$ = 0.13, 0.15, 0.19 and 0.23 were obtained under different growth conditions, depending upon applied oxygen pressure and growth rate. Nearly stoichiometric $SrFeO_{3.00}$ and slightly oxygen-deficient $SrFeO_{2.95}$ were prepared by post-annealing at 400°C under high oxygen pressure of 5 kbar and 700 - 800 bar, respectively. A more strongly oxygen-deficient sample of composition $SrFeO_{2.69}$ was obtained by post-annealing at 550°C in argon flow. In the following the samples will be labeled according to the oxygen contents determined by TG. An independent check of oxygen stoichiometries as well as an estimate of the phase compositions was derived from Mössbauer spectra (Section IV.A).

The magneto-transport characterization was performed with a PPMS system (Quantum Design). For measurements of the electrical resistivity, $\rho$, the crystals were cut and oriented along (100) faces, Cr/Au electrodes were deposited in a four-point contact geometry, and gold leads were glued on the electrodes with silver epoxy. The susceptibility curves were recorded in Field Cooling (FC) and subsequent Field Heating (FH) runs, with a 1T field. The electrical resistivities were measured in ac-mode in Zero Field Cooling (ZFC) and subsequent Zero Field Heating (ZFH), this was followed by FC and FH runs at 9T. The zero-field and in-field runs were compared to obtain the MR. We adopt the following definition: $MR = \frac{(\rho(H)-\rho(0))}{\rho(0)}$. Isothermal field scans were recorded near the temperature of the anomalies observed in FC and FH runs. For this series of measurements, the sample was first heated to room temperature and subsequently cooled to the required temperature in zero field.

Mössbauer spectra were measured with a standard spectrometer operating with a sine-type drive signal. The source was $^{57}Co$ in a Rh matrix. For the Mössbauer investigations, crystals of the different $SrFeO_{3-\delta}$ compositions were ground, and the resulting powder was diluted with polyethylene in order to ensure a homogeneous distribution of the material in the plexiglass sample container. The container was placed in an Oxford flow cryostat. The spectra normally were evaluated by fitting Lorentzians to the data. Spectra of magnetically ordered phases were evaluated with six-line patterns. Spectra of magnetically ordered $SrFeO_3$ were evaluated with the Voigt based fitting algorithm of the Mössbauer software package "RECOIL" [36]. The intensity ratios were taken as 3:2:1:1:2:3 (thin absorber approximation). All spectra are referenced to α-iron.

The Raman spectra were collected with incident light of wavelength 514.5 nm emitted by mixed gas Ar/Kr lasers on two triple-monochromator DILOR spectrometers. The spectrometers were equipped with identical charge coupled device (CCD) detectors cooled with liquid $N_2$. The spectra were calibrated at each temperature using the lines of an argon lamp. The samples $SrFeO_{3.00}$ and $SrFeO_{2.85}$ were studied in macro-Raman mode, while $SrFeO_{2.69}$ was studied in micro-Raman mode. Great care was taken to focus on the same point in the whole temperature range. All samples investigated were cut along the <001> directions and they had a mirror-like finish.

The ellipsometric measurements were performed with a home-built ellipsometer in combination with a fast-Fourier-transform interferometer at the infrared beamline of the ANKA synchrotron at Forschungszentrum Karlsruhe, Germany [37].

## IV. RESULTS AND DISCUSSION

### IV.A Mössbauer spectroscopy

#### Room temperature spectra

In Fig. 2 room temperature Mössbauer spectra of various $SrFeO_{3-\delta}$ samples studied in this work are shown. By comparison with literature data [31,32], the spectra can be used to identify the different phases present. The spectrum of the 5 kbar annealed sample essentially consists of a single line with an isomer shift IS which is typical for stoichiometric cubic $SrFeO_3$ (C phase). On the other hand, the spectrum of $SrFeO_{2.95}$ reveals an additional quadrupole doublet with a higher IS. This indicates the presence of the tetragonal (T) phase $Sr_8Fe_8O_{23}$ and can be assigned to an average charge $Fe^{3.5+}$ signal arising from the structurally distorted Fe2 sites. The spectrum of $SrFeO_{2.87}$ mainly reflects the features of the T phase and consists of the $Fe^{3.5+}$ quadrupole doublet in addition to an $Fe^{4+}$ single line. As the low-temperature spectra reveal also some C phase (see below), the $Fe^{4+}$ single line in the spectrum of $SrFeO_{2.87}$ is attributed to a superposition of the Fe1 and Fe3 sites of the T phase and $Fe^{4+}$ sites of cubic $SrFeO_3$. An additional $Fe^{3+}$ doublet with a much larger quadrupole splitting in the spectrum of $SrFeO_{2.81}$ is a signature of the structurally distorted $Fe^{3+}$ (Fe2) sites in the crystal structure of orthorhombic (O) $Sr_4Fe_4O_{11}$. Accordingly, $SrFeO_{2.81}$ corresponds to a mixture of the T phase as majority and the O phase as minority components. As expected, the sample with composition $SrFeO_{2.69}$ contains the O phase as its main constituent. This is again apparent from the characteristic Mössbauer spectrum. In addition to the $Fe^{3+}$ quadrupole doublet, a second doublet with a smaller quadrupole splitting is apparent. Based on its IS value, it is assigned to the square pyramidal $Fe^{4+}$ (Fe1) sites in the crystal structure of the O phase. Two further weak lines in the spectrum of $SrFeO_{2.69}$ are the inner components of sextets arising from magnetically ordered $Sr_2Fe_2O_5$. These additional lines are more readily seen in spectra measured over an extended range of Doppler velocities. Quantitative estimates of oxygen stoichiometries and phase fractions as well as details about magnetic and charge ordering phenomena are obtained from the temperature dependence of the Mössbauer spectra of the various materials, described below. Selected Mössbauer parameters and the sample compositions derived from data are given in Table 2.

**$SrFeO_{3.00}$**

From the absence of any $Fe^{3+}$ or $Fe^{3.5+}$ signals in the low and room temperature Mössbauer spectra of 5kbar annealed $SrFeO_{3.00}$ we had concluded that this sample is stoichiometric ($\delta < 0.02$) [29]. In order to clarify the origin of anomalies in the magnetic and magnetotransport data near 55 K (see Ref. [29] and Section IV.B), we have now performed a detailed Mössbauer study of the $SrFeO_{3.00}$ sample in the temperature range 10 to 130 K. Considering the pronounced hysteresis effects in the magnetotransport data, Mössbauer experiments have been performed in the heating as well as in the cooling mode. A selection of spectra is shown in Fig. 3. As expected, the spectra confirm the development of magnetic ordering below 130 K, indicated by six-line hyperfine patterns. Similar to the 10 K spectrum shown in Ref. [29], all spectra at $T \leq 60$ K consist only of a single sextet, with isomer shifts IS and magnetic hyperfine fields $B_{hf}$ in good agreement with earlier Mössbauer data of $SrFeO_3$ [31]. Within the error limits (about 2%) there is no indication of additional Fe sites in these spectra. Up to 60 K no major changes are seen. However, at somewhat higher temperature between 60 and 70 K some intensity near v = 0 mm $s^{-1}$ starts to emerge (v is the velocity of the Mössbauer drive), and between 70 and 90 K a certain fraction of a paramagnetic signal becomes apparent. Furthermore, above 60 K the outer lines of the magnetic sextet become asymmetric, and the lines broaden progressively up to 125 K. At 130 K, a single line spectrum is seen, which means that the Néel temperature for the present sample lies between 125 and 130 K. This is somewhat lower than the literature value $T_N = 134$ K for cubic $SrFeO_3$ [9]. At first glance the spectra of the paramagnetic phase consist of a single line. However, detailed inspection of the room temperature spectrum (Fig. 2) indicates a weak shoulder towards higher Doppler velocities.

Considering the Mössbauer spectra of $SrFeO_{2.87}$ and $SrFeO_{2.95}$ (see below), it appears likely that the occurrence of paramagnetic sites between 70 and 90 K as well as the small asymmetry in the room temperature spectrum indicate the presence of small regions of tetragonal $SrFeO_{3-\delta}$ even in the 5kbar annealed sample. In fact, the room temperature spectrum (Fig. 2) can be fitted by one $Fe^{4+}$ single line and a 4% contribution of $Fe^{3.5+}$ sites. The Mössbauer parameters of the latter sites were taken from a sample with increased oxygen deficiency, where the presence of the T phase is more obvious. Using the common oxidation states +2 and –2 for the Sr and O atoms, respectively, one obtains a sample composition $SrFeO_{2.99}$ which is within the error limit ±0.02 for δ derived from the TG data. At low temperatures the $Fe^{3.5+}$ sites in the tetragonal domains are expected to reveal charge localization (see below). The resulting 2% contribution of magnetically ordered $Fe^{3+}$ sites is, however, at the detection limit of the low temperature Mössbauer spectra.

It is apparent that the inner lines of the hyperfine sextets are narrower than the outer ones. This suggests that the line broadening above 60 K reflects a distribution of magnetic hyperfine fields. Accordingly the spectra were evaluated with a Voigt-based fitting algorithm [36]. In order to account for the asymmetry of the line shapes, a $B_{hf}$ distribution with two components was required: a major sharper and a minor broader one. The $B_{hf}$ distributions at selected temperatures are shown in Fig. 4. In the spectra at $T \geq 70$ K one $Fe^{3.5+}$ quadrupole doublet and one $Fe^{4+}$ singlet were added to account for the paramagnetic fraction. This model yields a reliable description of the spectral shapes (see Fig. 3). The area fraction of the magnetically ordered $Fe^{4+}$ sites (Fig. 5) decreases between 60 and 90 K, and changes only slightly upon further heating to 125 K. This means that the minority phase undergoes magnetic ordering near 70 K, which is similar to the Néel temperature of tetragonal $SrFeO_{3-\delta}$. The area fraction of the paramagnetic phase at 110 K is ~6% and is composed of 3% $Fe^{3.5+}$ and 3% $Fe^{4+}$ sites. These values are in reasonable agreement with the area fractions at room temperature. It is noted that the area fractions from spectra measured in the heating mode compare well with those from spectra measured in the cooling mode. Accordingly, there is no evidence of a hysteresis in the Mössbauer data. In Fig. 6 the temperature dependence of the mean values $\langle B_{hf} \rangle$ for the two components in the hyperfine field distribution is shown. The minor broad component 2 reveals a more pronounced temperature dependence than the major sharper component 1 and accounts for the increasing asymmetry of the spectra. The temperature dependence $\langle B_{hf} \rangle(T)$ of component 1 was reproduced by the relation $\langle B_{hf}(T) \rangle / \langle B_{hf}(0) \rangle = (1-T/T_N)^{\beta}$ with $\langle B_{hf}(0) \rangle = 33.5$ T, $T_N = 126.9$ K, $\beta = 0.16$.

In summary, the detailed Mössbauer study of the 5 kbar annealed sample $SrFeO_{3.00}$ indicates the presence of a small amount of oxygen vacancies ($\delta \sim 0.01$) which are concentrated in more strongly oxygen-deficient domains with a behavior similar to that of bulk tetragonal $SrFeO_{3-\delta}$. These domains are embedded in the majority cubic perovskite matrix. In the light of these results the question arises whether the 55 K transition in the magnetic and magnetotransport data (see [29] and Sect. IV.B) is an intrinsic property of pure cubic $SrFeO_3$, or rather related to phase inhomogeneity on a submicroscopic level. The transition in the magnetic and magnetotransport data occurs at somewhat lower temperature than the minority magnetic phase transition in the Mössbauer spectra. In addition there is no evidence for hysteresis effects in the Mössbauer area fractions whereas a quite pronounced hysteresis is observed in the magnetotransport anomaly. These findings suggest that the two transitions are of different origin. This conclusion is also supported by the Mössbauer and magnetotransport data of $SrFeO_{2.95}$, a sample with increased fraction of T phase (see below). Finally, we point out that the magnetic susceptibility of a $SrFeO_3$ sample annealed under even higher oxygen pressure is nearly identical to that of the sample annealed under 5 kbar [38]. In particular, the anomaly near 60 K is undiminished while the paramagnetic signals in the Mössbauer spectra between 70 and 130 K are absent. This underscores our conclusion that it is an intrinsic feature of the cubic phase.

**$SrFeO_{2.87}$**

In order to establish the properties of the T phase, we now turn to the Mössbauer spectra of $SrFeO_{2.87}$. The spectra shown in Fig. 7 provide evidence of two magnetic phase transitions, in agreement with magnetic susceptibility data of the same sample. The magnetic susceptibility curve of the $SrFeO_{2.87(2)}$ crystal is very similar to that of the $SrFeO_{2.85(2)}$ sample used in magnetotransport and optical experiments (Sections IV.B and IV.C below). The first transition below $T_{N1}$~130 K leads to a single hyperfine sextet D (area fraction ~20%), with isomer shifts IS and a temperature dependence of the magnetic hyperfine field $B_{hf}$ (T) (Fig. 9) typical for cubic $SrFeO_3$. Near $T_{N2}$ ~ 70 K, a second magnetic phase transition occurs. The complicated but well-structured low-temperature spectra were modeled by a superposition of five (15 K) or six (>15 K) hyperfine sextets. A typical decomposition is illustrated in Fig. 8 for 50 K, and the results of the data evaluation are depicted in Fig. 9. From $B_{hf}$ (T) it is evident that the magnetic order for all components except D vanishes near 70 K. Thus all sites except D are considered as intrinsic to the T phase. Based on the IS and $B_{hf}$ values, charge states of 3+ , 4+, and 3.5+ are assigned to the sub-spectra (A,B), (D,E,F), and C, respectively. Considering the low-temperature area fractions and assuming equal Debye-Waller factors for the various Fe sites, this leads to an average iron oxidation state of 3.74 and an oxygen content of 2.87 ($\delta$ = 0.13), in agreement with the TG data. Above 130 K, $SrFeO_{2.87}$ is completely in the paramagnetic state. Following previous work [31,32], the spectra were analyzed in terms of two components only: one $Fe^{4+}$ single line, and one average charge $Fe^{3.5+}$ quadrupole doublet. The area fraction of the $Fe^{3.5+}$ sites decreases from ~55% in the paramagnetic phase to a residual fraction of about 10% in the magnetically ordered phase. This suggests that the magnetic ordering near 70 K coincides with charge ordering (CO) of most of the $Fe^{3.5+}$ into $Fe^{3+}$ and $Fe^{4+}$ sites. A charge disproportionation of $Fe^{4+}$, as for instance in $CaFeO_3$ [16], should lead to higher $Fe^{3+}$ area fractions and is therefore excluded.

Based on the Mössbauer spectra and previously reported room temperature structural data [28], the following microscopic model for the physical properties of tetragonal $Sr_8Fe_8O_{23}$ is proposed. The high electrical conductivity in the paramagnetic phase arises from a delocalized electronic system with higher electron density at the Fe2 sites ($Fe^{3.5+}$) and lower electron density at the Fe1 and Fe3 sites (both formally $Fe^{4+}$). Although the coordination numbers are different, the two $Fe^{4+}$ species cannot be resolved in the Mössbauer spectra of the paramagnetic phase. Near 70 K the $Fe2^{3.5+}$ sites undergo a charge ordering transition, which leads to electronically localized $Fe^{3+}$ and $Fe^{4+}$ sites and to a concomitant development of three-dimensional magnetic order. Sextet F in the spectra can then be attributed to the $Fe^{4+}$ sites arising from the CO. Their more localized behavior is in agreement with a more negative IS than that in $SrFeO_3$ (c.f. with IS of sextet D) [39]. This view is supported by the nearly equal values of the total $Fe^{3+}$ area fraction due to sites A, B and of the area fraction of the $Fe^{4+}$ site F below $T_N$. The sum of the area fractions of the two other $Fe^{4+}$ sites D and E is comparable to the $Fe^{4+}$ fraction in the paramagnetic phase, but below $T_{N2}$ they are resolved due to their different $B_{hf}$. It is suggested that component E corresponds to magnetically ordered square pyramidal $Fe^{4+}$ sites in $Sr_8Fe_8O_{23}$ which are connected only to Fe2 atoms. In the course of the CO involving Fe2, the electrons at Fe1 also become more localized, which again gives rise to a more negative IS than in $SrFeO_3$. It is observed that the area fraction of sextet D decreases considerably between 70 and 90 K (see Fig. 9). This implies that a certain fraction of about 10 - 12% of D in the low temperature spectra is not due to cubic $SrFeO_3$, but intrinsic to the T phase. These sites possibly originate from the nearly regular $Fe(3)O_6$ octahedra.

The arguments considered so far rely on ideal stoichiometric $Sr_8Fe_8O_{23}$ corresponding to $SrFeO_{2.875}$. However, from the Mössbauer spectra it is evident that the present crystals of nominal composition $SrFeO_{2.87}$ contain a fraction of ~ 20% $Fe^{4+}$ sites arising from cubic $SrFeO_3$. This implies that an additional oxygen deficiency must be accommodated in the T phase compared to stoichiometric $Sr_8Fe_8O_{23}$. This is confirmed by the analysis of the 110 K spectrum where the C phase contribution is already magnetically ordered. The paramagnetic sub-spectrum that completely reflects the T phase is composed of 59% $Fe^{3.5+}$ and 41% $Fe^{4+}$ sites. For stoichiometric $Sr_8Fe_8O_{23}$ one would expect equal fractions of $Fe^{4+}$ and $Fe^{3.5+}$ sites. Accordingly there is an excess of $Fe^{3.5+}$ species, in agreement with the residual fraction of magnetically ordered $Fe^{3.5+}$ sites in the low temperature spectra. The latter sites are obviously not involved in CO and possibly are located at a crystallographically different site in the low temperature crystal structure. The above area fractions in the 110 K spectrum lead to the composition $SrFeO_{2.85}$ for the T phase in the present sample. There is no evidence of the orthorhombic phase in the spectra (see below). Finally it is noted that two magnetically distinct $Fe^{3+}$ sites are discernible, one of which (indicated by an arrow in Fig. 7) exhibits an unusual temperature dependence of $B_{hf}$. This behavior could also be related to disorder effects introduced by additional oxygen vacancies. Clearly the chemical, structural, electronic and magnetic behavior of the T phase is very complex, and more studies on crystals with improved phase homogeneity are required.

**$SrFeO_{2.95}$**

The 800 bar oxygen-annealed $SrFeO_{2.95}$ crystals are also a phase mixture of C and T phases, but with the C phase as the majority component. This is confirmed by the Mössbauer spectra shown in Fig. 10. At 100 K, for instance, the spectra show the coexistence of paramagnetic sites and the magnetically ordered C phase. From the area ratios, the fractions of C and T phases are estimated as 60% and 40%, respectively. Near 80 K the minority phase begins to develop magnetic order, and the low temperature spectra reveal the various Fe sites present in the charge-ordered T phase. The spectra are well described as a superposition of the spectra of C and T phases. These results also support our interpretation of the Mössbauer spectra of the 5 kbar annealed $SrFeO_3$, where the small minority component behaves in a fashion very similar to that of the T phase fraction in $SrFeO_{2.95}$. The temperature evolution of the present $SrFeO_{2.95}$ Mössbauer spectra is comparable to that of a ceramic sample of similar composition reported by Zhao et al. [27]. These authors attributed the paramagnetic signal near 80 K to a single $Fe^{(3+\delta)+}$ site, and included a single $Fe^{3+}$ line in addition to the $Fe^{4+}$ sextet in their model of the 4 K spectrum. In the light of the present work, it appears likely that the ceramic sample of Ref. [27] is also a mixture of C and T phases, and that the changes seen in the spectra near 80 K correspond to the magnetic/CO transition of the T phase. However, the signal-to-noise ratio of the Mössbauer spectra in Ref. [27] is not sufficient to resolve the complex low temperature sub-spectrum of the T phase.

**$SrFeO_{2.69}$ and $SrFeO_{2.81}$**

The O phase $Sr_4Fe_4O_{11}$ is characterized by magnetically disordered $Fe^{4+}$ sites down to 4 K [31,41]. The latter are indeed evident in the 4 K spectrum of the present $SrFeO_{2.69}$ sample (Fig. 11) which reveals an $Fe^{4+}$ quadrupole doublet in addition to magnetic hyperfine sextets. The $Fe^{4+}$ ions have been associated with the square pyramidal $Fe1O_5$ sites in the room temperature crystal structure of $Sr_4Fe_4O_{11}$ [28]. By contrast, the Fe2 sites with $Fe^{3+}$ ions located in chains of vertex-sharing $FeO_6$ octahedra undergo magnetic ordering near 230 K [42,33]. In the spectra at $T \leq 200$ K shown in Fig. 11, they give rise to one $Fe^{3+}$ hyperfine sextet with a large quadrupole interaction parameter. At $T \geq 250$ K, only a quadrupole doublet with large quadrupole splitting is observed (Table 2). The present spectra confirm that in $Sr_4Fe_4O_{11}$ only the $Fe^{3+}$ sublattice is magnetically ordered below $T_N$. The unusual coexistence of magnetically ordered and magnetically disordered sublattices at low temperature in $Sr_4Fe_4O_{11}$ has been attributred to magnetic frustration [41,28,33].

As expected from the oxygen content, the crystals of nominal composition $SrFeO_{2.69}$ are not single phase. The spectra at 250 K and at room temperature still reveal two magnetically ordered $Fe^{3+}$ sites with somewhat different IS and $B_{hf}$ values. They can be assigned to the octahedral and tetrahedral $Fe^{3+}$ sites of the magnetically ordered brownmillerite-type ferrite $Sr_2Fe_2O_5$. Considering the 250 K spectrum, where the O phase is completely paramagnetic, one estimates that 64% of O phase and 36% of $SrFeO_{2.50}$ are present. The average iron oxidation state derived from the data is 3.32, leading to an overall sample composition of $SrFeO_{2.66}$. This estimate is in reasonable agreement with the TG data.

Finally, the presence of the O phase as minority component in $SrFeO_{2.81}$ crystals (showing low temperature positive MR effects, see below) is confirmed by a magnetic $Fe^{3+}$ sextet with large quadrupole interaction in the 140 K and magnetically disordered $Fe^{4+}$ sites in the 12 K Mössbauer spectra (Fig. 12). The essential features of the spectra are well modeled by a superposition of those of the T and O phases. The fraction of the O phase is most reasonably estimated from the 140 K spectrum. As the $Fe^{3+}$ and $Fe^{4+}$ sites in the O phase occur in the ratio 1:1, its volume fraction is given by twice the area fraction of the magnetically ordered $Fe^{3+}$ component and amounts to 24%. The average iron oxidation state is 3.67, which leads to the overall composition $SrFeO_{2.84}$. Subtracting the $Fe^{4+}$ fraction of the O phase from the total $Fe^{4+}$ content, the composition of the T phase in the sample is derived as $SrFeO_{2.86}$. Again, there is an excess of $Fe^{3.5+}$ sites in comparison with stoichiometric $Sr_8Fe_8O_{23}$.

## IV.B Magnetic susceptibility and ac transport

### Magnetic susceptibility

Fig. 13 shows the magnetic susceptibility $\chi$ of a set of single crystals measured at 1 T in field cooling and heating runs. The temperatures corresponding to anomalies in the $\chi(T)$ curves are summarized in Table 3. In the $\delta = 0$ sample, a maximum heralds the onset of helical order at 130 K, and an additional pronounced anomaly of the susceptibility is observed at ~ 60 K (FC data). The latter anomaly shows a thermal hysteresis with a width of 10 K. In the $\delta = 0.05$ sample the maximum at 130 K is predominant, but a pronounced shoulder is observed around 75 K. This is followed by a decrease of the susceptibility upon further cooling. The latter feature is a signature of the onset of antiferromagnetic order in the minority tetragonal phase and also displays thermal hysteresis. In agreement with this interpretation, the $\delta = 0.15$ sample exhibits a sharp maximum at 70 K with a hysteresis of ~ 3

K width. The weaker shoulder near 130 K is evidence of residual C phase. A noteworthy additional anomaly of the susceptibility is observed at 115 K for the three compositions containing metallic C phase. This feature has no obvious counterpart in the Mössbauer spectra, and its microscopic origin is presently unknown. With a further increase of the oxygen deficit, the 70 K feature is broadened for $\delta$ = 0.19 and shifted towards lower temperature in the $\delta$ = 0.23 sample. In addition, the small inflection point in the susceptibility near 230 K and the slight upturn below 50 K are consistent with what is known of the orthorhombic phase [33]. The former feature reflects the antiferromagnetic ordering of the $Fe^{3+}$ sites near 230 K, whereas the latter is associated with the $Fe^{4+}$ sites which remain magnetically disordered. It is emphasized that the 230 K transition in $SrFeO_{3-\delta}$ samples corresponds to the well-known magnetic phase transition of the orthorhombic phase. It is *not* a new and unexpected feature as has been claimed recently [43].

**Magnetoresistance**

Figures 14 and 15 summarize the magnetoresistance (MR) properties of the $SrFeO_{3-\delta}$ system. The temperatures corresponding to anomalies in the MR are also included in Table 3. In agreement with earlier work [9], the temperature dependence of the resistivity of $SrFeO_{3.00}$ (Fig. 14a) indicates that the C phase is metallic. A precipitous decrease of the resistivity by a factor of two is observed at 52 K (ZF data). The transition exhibits a hysteresis with a width of 9 K and appears to be related to the 60 K anomaly in the magnetic susceptibility data. Application of a 9 T magnetic field shifts the transition by ~5 K to *higher* temperature. This induces a large negative MR effect (Fig. 15) in a narrow temperature range around 55 K ($\Delta T_{1/2}$ ~ 8 K, max. MR ~ -25%).

$SrFeO_{2.95}$ also exhibits metallic behavior with a resistivity anomaly below 70 K (Fig. 14b), which coincides with the 70 K maximum in the susceptibility of this sample. In contrast to the $SrFeO_{3.00}$ data, the resistivity first exhibits an upturn upon cooling below 70 K, followed by a decrease at lower temperature. Application of a 9 T magnetic field reveals that the resistivity anomaly in $SrFeO_{2.95}$ actually reflects two phase transitions. The shape of the $\rho(T)$ curve can be well understood as the superposition of the properties of C and T phases for this composition. This is evident by considering the magnetotransport data of $SrFeO_{2.85}$, where the T phase is the majority component. As opposed to the samples dominated by the C phase, the oxygen-deficient $SrFeO_{2.85}$ composition displays a weakly activated, semi-conducting behavior upon cooling from room temperature to 70 K, where the resistivity *increases* by an order of magnitude (Fig. 14c)). The resistivity jump is in agreement with a charge ordering transition near 70 K, as suggested by the Mössbauer spectra of $SrFeO_{2.87}$. The transition is characteristic of the T phase and shifted by ~6 K to *lower* temperature in a magnetic field of 9 T, in contrast to the 52 K transition in cubic $SrFeO_3$. The field-induced low temperature shift of the transition gives rise to a giant negative MR (GMR) effect of -90% that peaks sharply at ~69 K in a narrow temperature range ($\Delta T_{1/2}$~3.8K) (Fig. 15). Below 50 K, the small down-turn of the resistivity could be assigned to percolation paths of the metallic minority phase within the insulating matrix.

Returning now to the $SrFeO_{2.95}$ magnetotransport data, the small resistivity upturn near 70 K, which is shifted to lower temperature at 9 T, can be assigned to the charge ordering transition in the minority T phase. On the other hand, the overall resistivity decrease, which is shifted to higher temperature at 9 T, reflects the resistivity anomaly in the majority C phase. These considerations are in full agreement with the Mössbauer data on $SrFeO_{2.95}$. It is noteworthy that at 9 T, in contrast to the zero-field properties, the resistivity decrease in the C phase occurs at *higher* temperature than the CO transition in the T phase. As the oxygen deficit is further increased, the amount of metallic phase decreases and accordingly the percolation paths vanish. This is exemplified for $SrFeO_{2.81}$, where the weakly activated semi-conducting behavior is evidenced down to 70 K, and a resistivity jump of an order of magnitude occurs akin to the one of $SrFeO_{2.85}$. Below 50 K, however, an additional strong up-turn in $\rho$ is noticed, and the sample becomes fully insulating at low temperatures. Finally, the resistivity of the $SrFeO_{2.77}$ samples exhibits behavior closely similar to that of $SrFeO_{2.81}$, but the jump of the resistivity shifts slightly to lower temperature and the thermal hysteresis is reduced. This presumably reflects the increasing volume fraction of the orthorhombic phase.

We now address the magnetic field dependence of the resistivity, which exhibits a variety of behaviors as a function temperature, oxygen deficit, and phase mixture. Representative data are shown in Fig. 16. The samples with the C phase as majority component ($\delta$=0.00 and $\delta$=0.05) have a large negative MR effect in a quite narrow temperature range (Fig. 15). Within this temperature range, the resistivity is weakly field dependent up to about 8 T, and is abruptly reduced for larger fields. Addition of a minority amount of tetragonal phase lowers the critical field significantly to ~5.5 T (Fig. 16a). For $\delta$=0.00 and $\delta$=0.05, the MR effect is remnant after a subsequent ramp-down of the field, and even when the field is reversed (data not presented here). For the samples containing predominantly the T phase (Fig. 15), a giant negative MR effect is observed. This effect also requires fields around 8 T (Fig. 16b), but it is fully reversible upon reducing the field back to zero, and a symmetric curve is observed when the field is reversed. Finally, a pronounced *positive* MR effect is apparent in the low temperature insulating phases of $SrFeO_{2.81}$ and $SrFeO_{2.77}$. For these compositions, the MR effect is history dependent, e.g. the value of MR measured at a given temperature (T = 10 K) changes between isothermal field scans and temperature scans. From the first method a MR effect of 52% is derived, while for the second a positive MR as large as 256% is measured.

The origin of the main MR effects in the C and T phases appears to be a shift in the thermodynamic balance of two phases with different conductivities, which induces a shift in the critical temperature separating these phases. A similar effect has been observed at the Verwey transition of magnetite [44]. The more complex behavior shown in Fig. 15, as well as the positive MR exhibited by the insulating samples at low temperatures, probably reflect at least in part the field dependence of current paths in a mixture of phases with different conductivities. If this is the case, a full understanding of these effects requires detailed information about the microstructure of these samples, an interesting subject of future work.

We now compare the present magnetotransport data on $SrFeO_{3-\delta}$ single crystals with the study by Zhao *et al.* [27] on ceramic samples. One of our crystals had the same composition, $SrFeO_{2.95}$, as the one investigated by Zhao *et al*. Considering the similar composition and the similar Mössbauer spectra of the two materials (see above) it is likely that the sample of Ref. [27] also contains a mixture of cubic and tetragonal phases. The pronounced $\rho(T)$ anomaly near 50 K in Ref. [27] and the associated MR effect then reflect the transition of the majority C phase, whereas the magnetic anomaly at a somewhat higher temperature (65 K) and the development of paramagnetic sites in the Mössbauer spectra correspond to the magnetic/CO transition in the minority T phase. In contrast to the present $SrFeO_{2.95}$ single crystal, the signature of the CO transition is not seen in the $\rho(T)$ data of the ceramic sample. However the detailed electronic transport paths are determined by the microstructure of the materials, which may differ in single crystals and ceramics. We emphasize that the pure C phase is metallic both below and above 55 K ruling out that the resistivity anomaly coincides with a metal – insulator transition. Accordingly the low temperature upturn in the resistivity data of Ref. [27] reflects the coexistence of metallic C and insulating T phases rather than a metal-insulator transition in $SrFeO_{2.95}$ as was claimed by Zhao *et al*. [45]. In the present single crystals, large MR effects are only found in the vicinity of the phase transitions in the respective compositions. The low temperature MR effects in ceramic $SrFeO_{2.95}$ possibly correspond to grain boundary effects in the polycrystalline material. A further weak resistivity anomaly coupled with a small MR effect was established near 107 K in ceramic $SrFeO_{2.95}$ and associated with the magnetic ordering transition. A similar feature near 115 K is also apparent in the present data of $SrFeO_{3.00}$. It is, however, rather related to an additional cusp near 115 K in the magnetic susceptibility (indicated by an arrow in Fig. 13), than to the onset of helical ordering near 130 K. The relation between magnetic and resistivity anomalies is clarified in Fig. 17 where we compare the derivatives $d\ln\chi/dT$ and $d\ln\rho/dT$ for $SrFeO_{3.00}$. It is obvious that the 130 K magnetic transition does not give rise to any anomaly in $\rho(T)$. Fig.17 also suggests that the resistivity response related to the 60 K magnetic transition occurs at somewhat lower temperature.

The negative GMR effect near 70 K in mainly tetragonal materials has recently been confirmed for a ceramic sample of $SrFeO_{3-\delta}$ with $\delta = 0.17$ [43]. This material appears to be a mixture of T and O phase, as can be inferred from the presence of a 230 K transition in the magnetic susceptibility data. The onset of the GMR effect is already seen at 6 T, a somewhat lower field than in the present $\delta = 0.15$ sample. Finally, we note that a combined magnetic ordering/CO transition giving rise to a similar resistivity jump as in $SrFeO_{2.85}$ was reported previously for $Sr_{2/3}Ln_{1/3}FeO_3$ [46]. In this system an antiferromagnetic phase transition is coupled to a charge disproportionation of $Fe^{4+}$, which results in $Fe^{3+}Fe^{3+}Fe^{5+}$ CO along the pseudocubic [111] direction [47]. For samples with Ln = La and Pr, a weak MR effect of -2% at 7 T was observed near $T_N$. Somewhat larger MR effects up to -7% were reported at low temperatures [48]. However, that there is no charge disproportionation of $Fe^{4+}$ in the charge-ordered state of $SrFeO_{2.85}$.

### IV.C Raman scattering and far-infrared ellipsometry

In the cubic perovskite $SrFeO_3$ (space group Pm-3m) no Raman vibrational modes are allowed, but three IR active modes of $F_{1u}$ symmetry are expected. This is in agreement with the experimental Raman and ellipsometric IR spectra of $SrFeO_{3.00}$ depicted in Fig. 18. While the Raman spectra do not show any features attributable to phonons, three IR modes with room temperature frequencies 172, 249, and 559 $cm^{-1}$ are observed. The temperature independence of the spectra corroborates the absence of any structural phase transitions below room temperature. The phonon modes are superimposed on an electronic background due to a Drude-like charge carrier response, which is strongly enhanced at low temperature. This compares well with the increased electrical conductivity below the 55 K phase transition.

Further evidence for the CO transition in the T phase is obtained from Raman and infrared spectra of $SrFeO_{2.85}$. Due to symmetry lowering, Raman modes become allowed for the T and O phase even in the absence of CO. In agreement with this expectation, the Raman spectra of the $\delta$=0.15 and $\delta$=0.31 samples, shown in Fig. 19, exhibit several phonon features at room temperature. For $SrFeO_{2.85}$ (Fig. 19, upper panel), several new vibrational modes appear upon cooling below 70 K, the CO temperature of the T phase already discussed in Sections IV.A and IV.B. As expected on general grounds, this shows that the CO transition is coincident with a structural transition. For the $SrFeO_{2.69}$ sample (Fig. 19, bottom panel), which does not contain any T phase, no new Raman-active phonons are observed upon cooling to low temperatures. This is in agreement with a recent neutron diffraction study on $Sr_4Fe_4O_{11}$, which has shown that the orthorhombic crystal structure is retained down to 1.5 K [33]. This implies that the orthorhombic phase does not undergo a CO instability.

These conclusions are supported by far-infrared ellipsometry data on the $\delta$=0.15 sample. It was reported already in our previous work [29] that the optical conductivity suddenly drops upon cooling below ~70 K, and that additional optical modes appear in the same temperature range. Figure 20 provides a synopsis of room temperature and low temperature Raman scattering data in two different polarization geometries as well as far-infrared ellipsometry data on this sample. Note that comparable changes in the optical conductivity have been reported for the CO transition in $Sr_{2/3}La_{1/3}FeO_3$ [49]. All data sets exhibit numerous additional phonon modes at low temperatures. Their frequencies are summarized in Table 4. Since the crystal structure of the tetragonal phase that undergoes the CO transition has thus far not been determined, it is impossible to assign the modes. However, it is evident that a number of modes are observed in both Raman polarization geometries and in the ellipsometry data at the same frequencies. This may either reflect the low symmetry of the crystal structure in the CO state, or a symmetry lowering at the boundaries of the constituent T and C phases.

## V CONCLUDING REMARKS

Iron(IV) rich oxides reveal a variety of magnetoresistance effects. In the previously studied cobalt or nickel substituted materials, the generation of MR is related to creation of ferromagnetic clusters or regions. The largest MR effects are encountered in compositions with spin-glass like magnetic behavior [22] pointing to competing ferro- and antiferromagnetic interactions. In this respect these materials behave in a similar fashion as other magnetoresistive oxides, for instance certain cobaltites [50]. In contrast, the MR effects in single crystals of $SrFeO_{3-\delta}$ which are the subject of the present work are apparently different as they do not involve ferromagnetic ordering. We have shown that it is essential to consider the phase composition of the samples in order to establish the physical properties of this system. Actually three different types of MR effects are found. In cubic $SrFeO_3$ helical magnetic ordering emerges near 130 K which is, however, not associated with any anomalies in the electrical resistivity. In contrast a hitherto unknown magnetic transition near 60 K gives rise to a sharp decrease of the resistivity. The latter is shifted by a magnetic field of 9 T to higher temperature which results in a large MR effect. The detailed nature of the 60 K transition remains to be established, but it obviously neither involves ferromagnetic ordering nor charge reorganization. A coincident charge – magnetic ordering transition is the origin of a resistivity jump by an order of magnitude in materials containing tetragonal oxygen-deficient $SrFeO_{3-\delta}$ (ideal composition $Sr_8Fe_8O_{23}$) as major component. The transition is shifted by a 9 T field to lower temperature which results in a giant negative MR effect of 90%. This is the largest MR effect discovered so far in an iron(IV) based oxide. The magnetic phase transition shows antiferromagnetic characteristics, but the detailed low-temperature spin and crystal structures remain to be clarified. Finally, there is a pronounced positive MR effect in low-temperature insulating $SrFeO_{3-\delta}$ materials which may either be a property of the pure tetragonal phase or related to the coexistence of tetragonal and orthorhombic phases.

We believe that exploration of the different types of magnetoresistance effects in iron(IV) based oxides is a valuable addition to our understanding of the magnetotransport properties of transition metal oxides and may broaden the concepts for the design of magnetoresistive materials and devices.

## Acknowledgment

We thank S. Ahlert, E. Brücher, K. Förderreuther, W. Hölle, Y.-L. Mathis, and F. Schartner for generous technical assistance, and L. Alff for valuable discussions.

**Figure Captions**

FIG. 1 (Color online) Illustration of the crystal structures of $Sr_8Fe_8O_{23}$ (top) and $Sr_4Fe_4O_{11}$ (bottom).

FIG. 2 (Color online) Room temperature Mössbauer spectra of ground $SrFeO_{3-\delta}$ crystals. Sub-spectra.: blue - $Fe^{4+}$, red - $Fe^{3.5+}$, green - $Fe^{3+}$ .

FIG. 3 (Color online) Selected Mössbauer spectra of 5kb oxygen-annealed $SrFeO_{3.00}$. The data were evaluated with a Voigt based fitting procedure as described in the text. The sub-spectra of the paramagnetic $Fe^{4+}$ (single line) and $Fe^{3.5+}$ (doublet) sites are drawn.

FIG. 4 (Color online) Magnetic hyperfine field distribution $P(B_{hf})$ at 90 K (right), 110 K (middle), and 125 K (left) obtained from the evaluation of the magnetic hyperfine patterns of $SrFeO_{3.00}$ by a Voigt-based fitting algorithm.

FIG. 5 (Color online) Temperature dependence of the area fraction of the magnetically ordered $Fe^{4+}$ sites in $SrFeO_3$ (blue: heating mode data, red: cooling mode data).

FIG. 6 (Color online) Temperature dependence of the mean values $\langle B_{hf}\rangle$ of the two hyperfine field components in the $B_{hf}$ distribution for $SrFeO_{3.00}$. The solid line corresponds to a fit of the relation $\langle B_{hf}\rangle(T)/ \langle B_{hf}\rangle(0) = (1-T/T_N)^{\beta}$ to the data of the main component.

FIG. 7 Mössbauer spectra of $SrFeO_{2.87}$ at selected temperatures. Solid lines correspond to the best fit to the data. See text for details. The arrow in the 50 K spectrum emphasizes the splitting of the $Fe^{3+}$ signal into two components.

FIG. 8 (Color online) Decomposition of the 50 K Mössbauer spectrum of $SrFeO_{2.87}$ into sub-spectra. See text for details.

FIG. 9 (Color online) Temperature dependence of isomer shifts (IS) (top), hyperfine fields $B_{hf}$ (middle), and area fractions (bottom) for the various iron sites of $SrFeO_{2.87}$. Full symbols correspond to magnetically ordered sites A – F, open symbols to the two paramagnetic components. In the area fraction plot the full circles correspond to the sum of the $Fe^{3+}$ sites A and B.

FIG. 10 Mössbauer spectra of $SrFeO_{2.95}$ at selected temperatures. The solid line corresponds to a best fit assuming a superposition of the spectra of the cubic and tetragonal phase.

FIG. 11 (Color online) Mössbauer spectra of $SrFeO_{2.69}$ at selected temperatures. The sub-spectra arising from the majority O phase are also drawn (blue : $Fe^{4+}$, red : $Fe^{3+}$ signal). The other lines correspond to the hyperfine sextets of the minority $Sr_2Fe_2O_5$ component.

FIG. 12 (Color online) Mössbauer spectra of $SrFeO_{2.81}$ at selected temperatures. The presence of the O phase is most clearly apparent from the $Fe^{3+}$ sextet with large quadrupole interaction (red sub-spectrum) and the magnetically disordered $Fe^{4+}$ (blue sub-spectrum) sites in the 12 K spectrum.

FIG. 13 Magnetic susceptibility of $SrFeO_{3.00}$, $SrFeO_{2.95}$, $SrFeO_{2.85}$, $SrFeO_{2.81}$ and $SrFeO_{2.77}$ single crystals, measured in field cooling and subsequent field heating runs at B=1T. The curves for the latter four samples were shifted by the amounts indicated in the legend.

FIG. 14 (Color online) Resistivity in zero field cooling and heating runs (black) compared to field cooling and heating runs with a 9 Tesla field (red) for a) $SrFeO_{3.00}$, b) $SrFeO_{2.95}$, c) $SrFeO_{2.85}$ d) $SrFeO_{2.81}$ and e) $SrFeO_{2.77}$. As the absolute value of the resistance measurements (lines) was influenced by micro-cracks, these data were normalized around room temperature to IR data extrapolated to zero frequency (see [29]).

Fig. 15 Temperature dependence of the magnetoresistance MR at 9 Tesla for the five compositions. MR values were calculated according to $MR(9T) = (\rho(9T)-\rho(0))/\rho(0)$ from the isothermal field scans at the temperatures indicated by dots.

FIG. 16 (Color online) Field dependence of the magnetoresistance $MR(B) = (\rho(B)-\rho(0))/\rho(0)$ for the five compositions as derived from isothermal field scans at the temperatures where the respective MR effects are most pronounced: a) large negative MR for $SrFeO_{3.00}$ at T = 58 K (in black) and for $SrFeO_{2.95}$ at T = 65 K (in red), b) giant negative MR effect for $SrFeO_{2.85}$ at T = 68 K and c) positive MR effect for $SrFeO_{2.81}$ (black) and $SrFeO_{2.77}$ (red), both at T = 10 K.

FIG. 17 (Color online) Derivatives dln$\chi$/dT (blue) and dln$\rho$/dT (black) of the magnetic susceptibilities and resistivities of $SrFeO_{3.00}$.

FIG. 18 (Color online) Ellipsometric far-infrared (a) and Raman spectra (b) of cubic $SrFeO_{3.00}$ at selected temperatures. In order to facilitate comparison offsets have been added to the Raman spectra.

FIG. 19 (Color online) Temperature dependence of the Raman spectra a) polarized Raman spectra of $SrFeO_{2.85}$ in ZZ and b) unpolarized spectra of $SrFeO_{2.69}$. The incident light is the green line of an Ar laser at $\lambda$ = 514.5 nm. The $SrFeO_{2.85}$ sample was cut along the <100> direction, $SrFeO_{2.69}$ was polycrystalline. The asterisk in panel b) stands for a plasma line. The dotted lines that extend in both panels are guides-to-the-eye to observe the shifts of some common modes.

FIG. 20 (Color online) Comparison of the Raman and far-infrared ellipsometry spectra for the charge-order composition $SrFeO_{2.85}$ at low temperature and room temperature a) Raman spectra in parallel (ZZ) and cross (XZ) polarization, b) far-infrared ellipsometry spectra: optical conductivity $\sigma$ and real part of the dielectric permittivity $\varepsilon_1$.

**Table 1** Iron-oxygen distances for the various members of the $Sr_nFe_nO_{3n-1}$ series of ferrates.

| Compound | Bond distances (Å) | Valence assignment | Rel. abundance |
|---|---|---|---|
| $SrFeO_3$ [28] | 6 x 1.926 | $Fe^{4+}$ | 1 |
| $Sr_8Fe_8O_{23}$[28] | **Fe1** 4 x 1.851, 1 x 1.926<br>**Fe2** 2x1.931, 2x1.952, 2 x 2.036<br>**Fe3** 4 x 1.912, 2 x 1.925 | $Fe^{4+}$<br>$Fe^{3.5+}$<br>$Fe^{4+}$ | 1:2:1 |
| $Sr_4Fe_4O_{11}$[28] | **Fe1** 4 x 1.855, 1 x 1.90<br>**Fe2** 4 x 2.044, 2 x 1.937 | $Fe^{4+}$<br>$Fe^{3+}$ | 1:1 |
| $Sr_2Fe_2O_5$[34] | **Fe1** 2 x 1.87, 2 x 2.091, 2 x 2.180<br>**Fe2** 2 x 1.916, 1 x 1.819, 1 x 1.980 | $Fe^{3+}$<br>$Fe^{3+}$ | 1:1 |

**Table 2** Selected Mössbauer parameters (IS: isomer shift, $B_{hf}$: hyperfine field, ε: quadrupole interaction parameter, $\Delta E_Q$: quadrupole splitting) derived from the evaluation of the data of $SrFeO_{3-\delta}$ samples. Errors are less or equal one in the last digit if not given explicitly in parenthesis.

| sample[1] | T (K) | Assignment | IS (mm $s^{-1}$) | $B_{hf}$ (Tesla) | ε or $\Delta E_Q$ (mm $s^{-1}$) | rel. Area (%) | derived composition |
|---|---|---|---|---|---|---|---|
| $SrFeO_{3.00}$ | 8 | $Fe^{4+}$ | 0.16 | 32.2 | - | 100 | |
| | 101 | $Fe^{4+}$ | 0.16 | 23.9, 19.7(8)[2] | - | 94.6(1.0) | |
| | | $Fe^{4+}$ | 0.14(3) | - | - | 2.5(8) | $SrFeO_{2.99}$ |
| | | $Fe^{3.5+}$ | 0.26[3] | - | 0.66[3] | 2.9(7) | 94.6% C + 5.4% T |
| | 299 | $Fe^{4+}$ | 0.07 | - | - | 96 | $SrFeO_{2.99}$ |
| | | $Fe^{3.5+}$ | 0.16[3] | | 0.6[3] | 4 | |
| $SrFeO_{2.95}$ | 7 | $Fe^{3+}$ | 0.41 | 45.1 | - | 12 | |
| | | $Fe^{4+}$ | 0.16 | 32.5 | - | 64 | |
| | | $Fe^{4+}$ | 0.03 | 23.8 | -0.09 | 11 | |
| | | $Fe^{4+}$ | -0.02 | 27.3 | 0.03(2) | 7 | |
| | | $Fe^{3.5+}$ | 0.27(2) | 38.9 | - | 5 | |
| | 100 | $Fe^{4+}$ | 0.16 | 26.3 | - | 60 | $SrFeO_{2.95}$ |
| | | $Fe^{3.5+}$ | 0.27 | - | - | 21 | 60% C + 40% T |
| | | $Fe^{4+}$ | 0.13 | - | - | 19 | |
| | 295 | $Fe^{4+}$ | 0.06 | - | - | 69 | |
| | | $Fe^{3.5+}$ | 0.16 | - | 0.56 | 31 | |
| $SrFeO_{2.87}$ | 15 | $Fe^{3+}$ | 0.41 | 45.0 | 0.02 | 18 | $SrFeO_{2.87}$ |
| | | $Fe^{4+}$ | 0.16 | 32.3 | - | 32 | |
| | | $Fe^{4+}$ | 0.04 | 23.9 | -0.08 | 22 | |
| | | $Fe^{4+}$ | -0.01 | 27.2 | 0.03 | 16 | |
| | | $Fe^{3.5+}$ | 0.23 | 39.3 | 0.00(2) | 12 | |
| | 110 | $Fe^{4+}$ | 0.16 | 24.2 | - | 19 | $SrFeO_{2.88}$ |
| | | $Fe^{3.5+}$ | 0.27 | - | 0.66 | 47(2) | 81% T + 19% C |
| | | $Fe^{4+}$ | 0.13 | - | - | 34(2) | T: $SrFeO_{2.85}$ |
| | 295 | $Fe^{4+}$ | 0.05 | - | - | 45 | |
| | | $Fe^{3.5+}$ | 0.16 | - | 0.60 | 55 | |
| $SrFeO_{2.81}$ | 12 | $Fe^{3+}$ | 0.42 | 44.8 | 0.02 | 20 | |
| | | $Fe^{4+}$ | 0.14(2) | 32.1 | -0.03(2) | 12 | |
| | | $Fe^{4+}$ | 0.04 | 23.8 | -0.10 | 21 | |
| | | $Fe^{4+}$ | 0.00 | 27.4 | 0.03 | 15 | |
| | | $Fe^{3.5+}$ | 0.25(2) | 39.6(2) | - | 10 | |
| | | $Fe^{3+}$ | 0.50(2) | 45.7 | -0.66 | 13 | |
| | | $Fe^{4+}$ | -0.04 | - | 0.35[3] | 8 | |
| | 140 | $Fe^{3+}$ | 0.45 | 38.6 | -0.67 | 12 | $SrFeO_{2.84}$, 24% O |
| | | $Fe^{3.5+}$ | 0.24 | - | 0.68 | 47 | 76% T + 24% O |
| | | $Fe^{4+}$ | 0.12 | - | - | 42 | T: $SrFeO_{2.86}$ |
| | 295 | $Fe^{4+}$ | 0.04 | - | - | 42 | |
| | | $Fe^{3.5+}$ | 0.15 | - | 0.64 | 48 | |
| | | $Fe^{3+}$ | 0.35 | - | 1.35 | 10 | |
| $SrFeO_{2.69}$ | 4 | $Fe^{3+}$(o)[4] | 0.51 | 52.6 | -0.35 | 18 | |
| | | $Fe^{3+}$(t)[4] | 0.29 | 45.3 | 0.31 | 20 | |
| | | $Fe^{3+}$ | 0.47 | 45.3 | -0.67 | 33 | |
| | | $Fe^{4+}$ | -0.03 | - | 0.35 | 29 | |
| | 250 | $Fe^{3+}$(o)[4] | 0.41 | 50.5 | -0.34 | 17 | $SrFeO_{2.66}$ |
| | | $Fe^{3+}$(t)[4] | 0.21 | 42.8 | 0.30 | 20 | 63% O + 37% B |
| | | $Fe^{3+}$ | 0.38 | - | 1.38 | 32 | |
| | | $Fe^{4+}$ | -0.03 | - | 0.32 | 32 | |

| sample[1] | T (K) | Assignment | IS (mm s$^{-1}$) | $B_{hf}$ (Tesla) | ε or $\Delta E_Q$ (mm s$^{-1}$) | rel. Area (%) | derived composition |
|---|---|---|---|---|---|---|---|
| | 295[5] | $Fe^{4+}$ | -0.09 | - | 0.32 | 48 | |
| | | $Fe^{3+}$ | 0.34 | - | 1.35 | 52 | |

[1] sample compositions estimated from TG data.
[2] Data evaluation with Voigt based fitting algorithm. See text for details.
[3] Parameter was not varied.
[4] Octahedral (o) and tetrahedral (t) sites from the brownmillerite phase $Sr_2Fe_2O_5$.
[5] Area fraction normalized to 100% O phase.

**Table 3** Summary of the temperatures $T_m$ and $T_\rho$ corresponding to anomalies in the magnetic susceptibility $\chi(T)$ and zero-field electrical resistivity $\rho(T)$ curves. The $T_m$ and $T_\rho$ values were obtained from the derivatives $d\ln\chi/dT$ and $d\ln\rho/dT$, respectively. The first value corresponds to the cooling mode, the values in parenthesis to the heating mode data.

| Compound | $T_m$(K) | $T_\rho$(K) |
|---|---|---|
| $SrFeO_{3.00}$ | 131.9 (132.0)<br>114.9 (116.1)<br>60.9 (70.1) | <br>110 (112)<br>52 (62) |
| $SrFeO_{2.95}$ | 131.8 (131.1)<br>115.8 (115.5)<br>69.8 (74.1) | <br>114 (115)<br>63 (73) |
| $SrFeO_{2.85}$ | 131.1 (131.1)<br>115.1 (115.1)<br>71.0 (74.1) | <br><br>70 (74) |
| $SrFeO_{2.81}$ | 50.9 (62.1)<br>229.4 (230.0) | 50 (46) (62) |
| $SrFeO_{2.77}$ | 47.9 (50.2, 62.2)<br>225 (226) | 48 (52) (60) |

**Table 4** Comparison of the phonon frequencies in far infrared ellipsometry and Raman scattering for mostly tetragonal $SrFeO_{2.85}$ sample at low temperature.

| | | | | | | | | | | | | |
|---|---|---|---|---|---|---|---|---|---|---|---|---|
| Raman ZZ ($cm^{-1}$)<br>Scattering XZ | 106<br>107 | | 140<br>139 | 159<br>160 | 178<br>177 | 194<br>195 | 205<br>204 | 242<br>244 | 258<br>260 | 268 | | 293 |
| Far-infrared ellipsometry ($cm^{-1}$) | 114 | 121 | 138 | 160 | 175 | 195 | 213 | 247 | 259 | 267 | 280 | 296 |
| Raman ZZ ($cm^{-1}$)<br>Scattering XZ | 303<br>304 | 323<br>322 | 337 | 355<br>354.5 | 391<br>391 | 413<br>415 | | 481<br>481 | 525<br>520 | | 567<br>571 | 616<br>612 |
| Far-infrared ellipsometry ($cm^{-1}$) | 308 | | | | 397 | | 427 | | | 539 | 570 | |

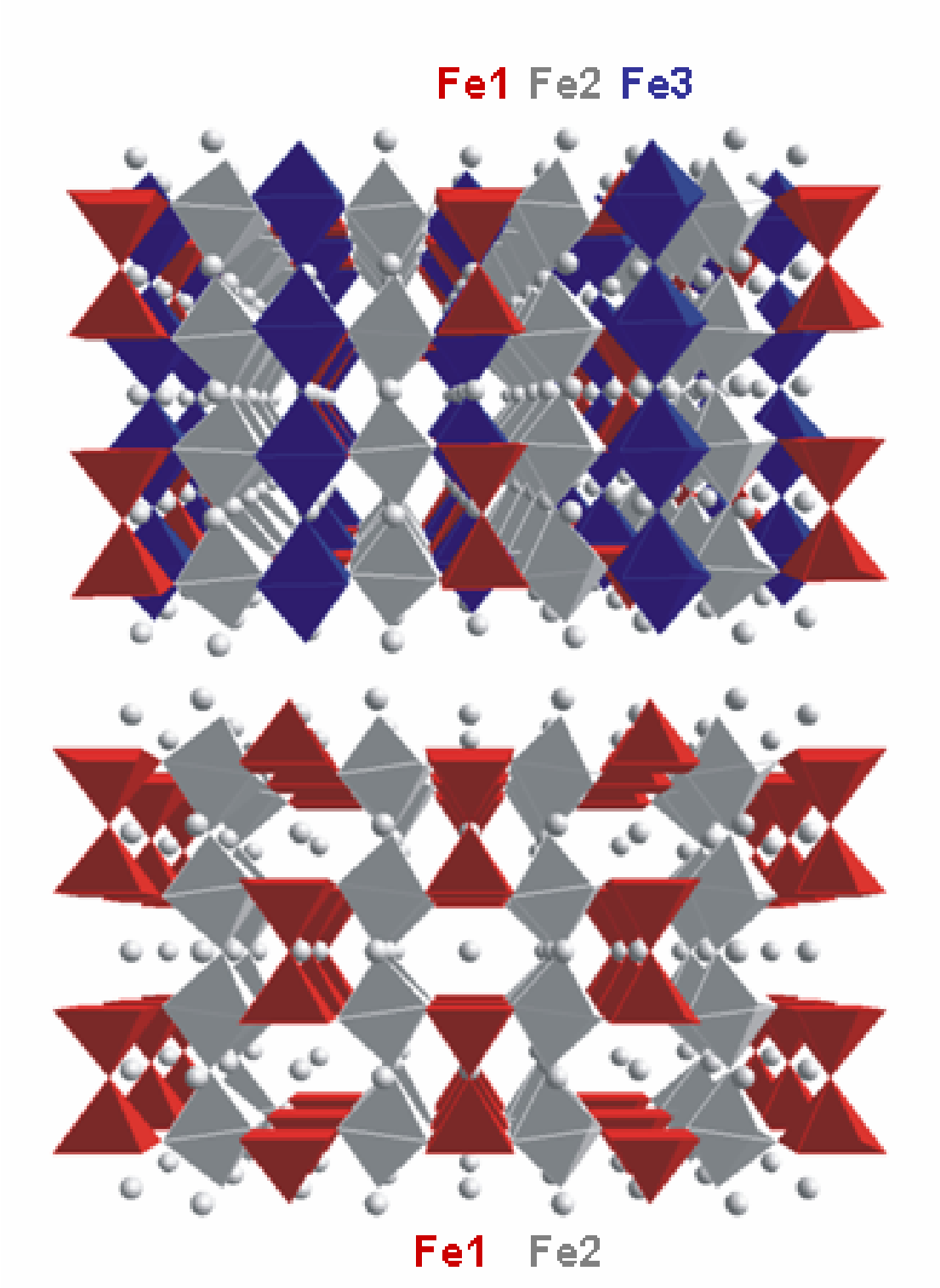

FIG.1

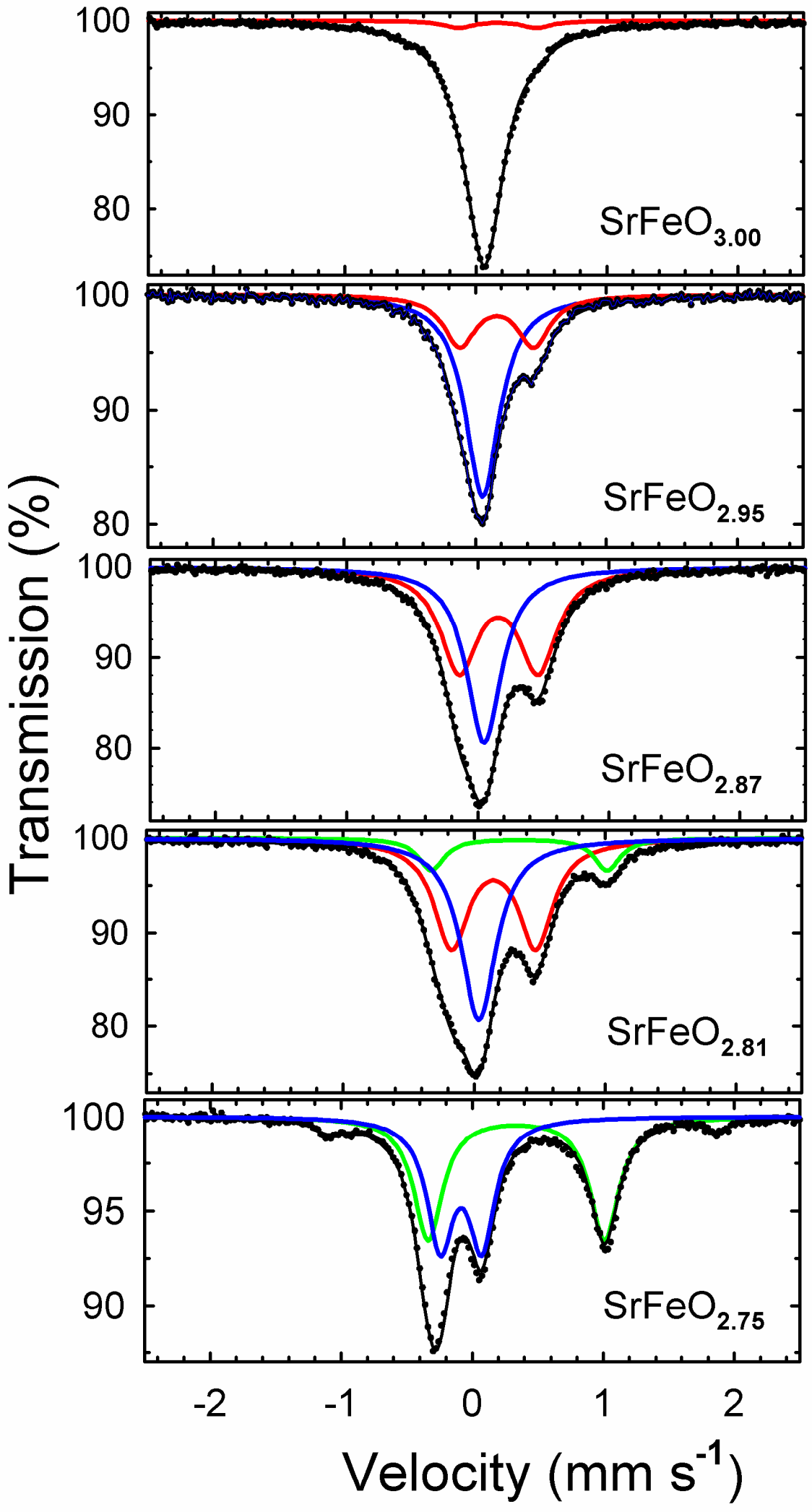

FIG.2

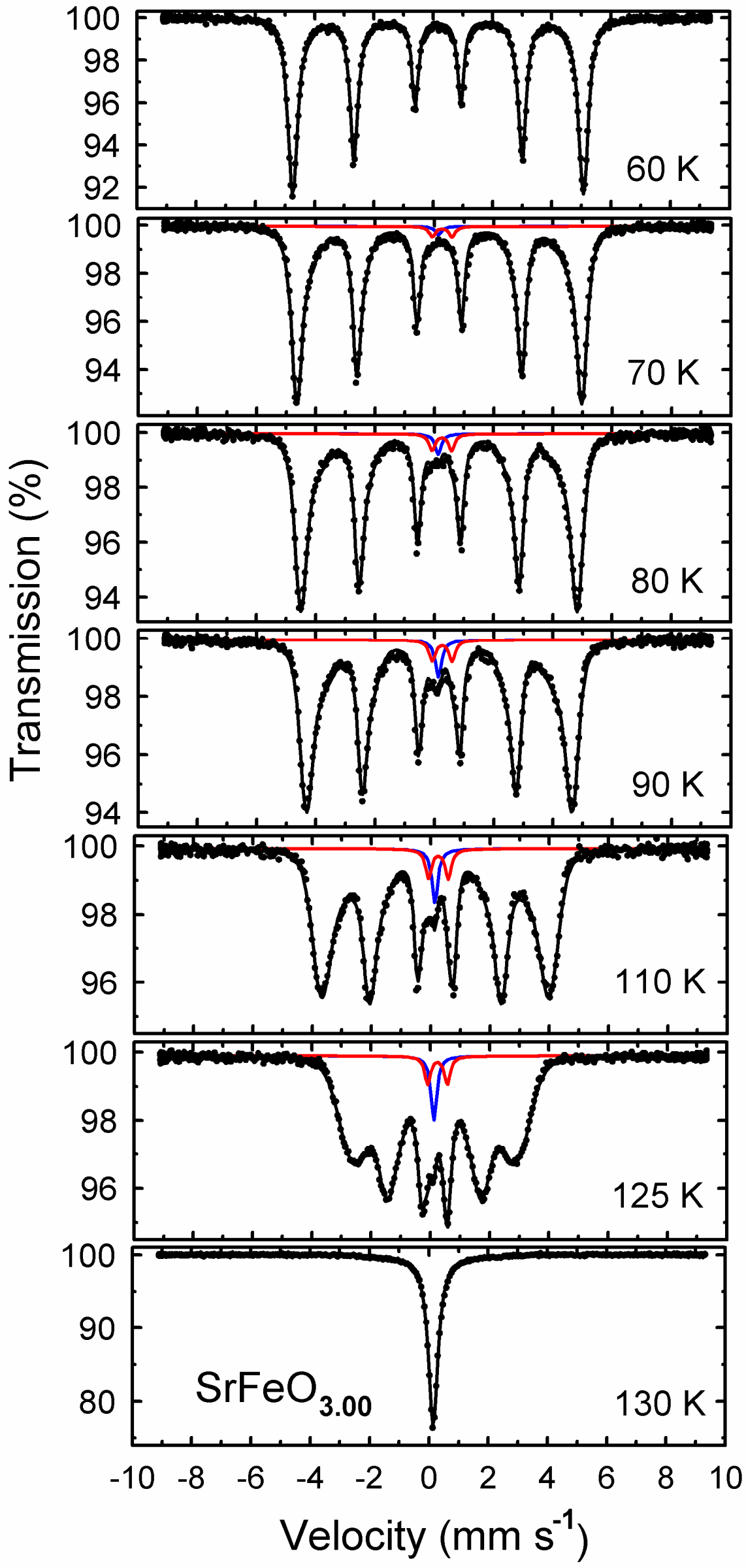

FIG.3

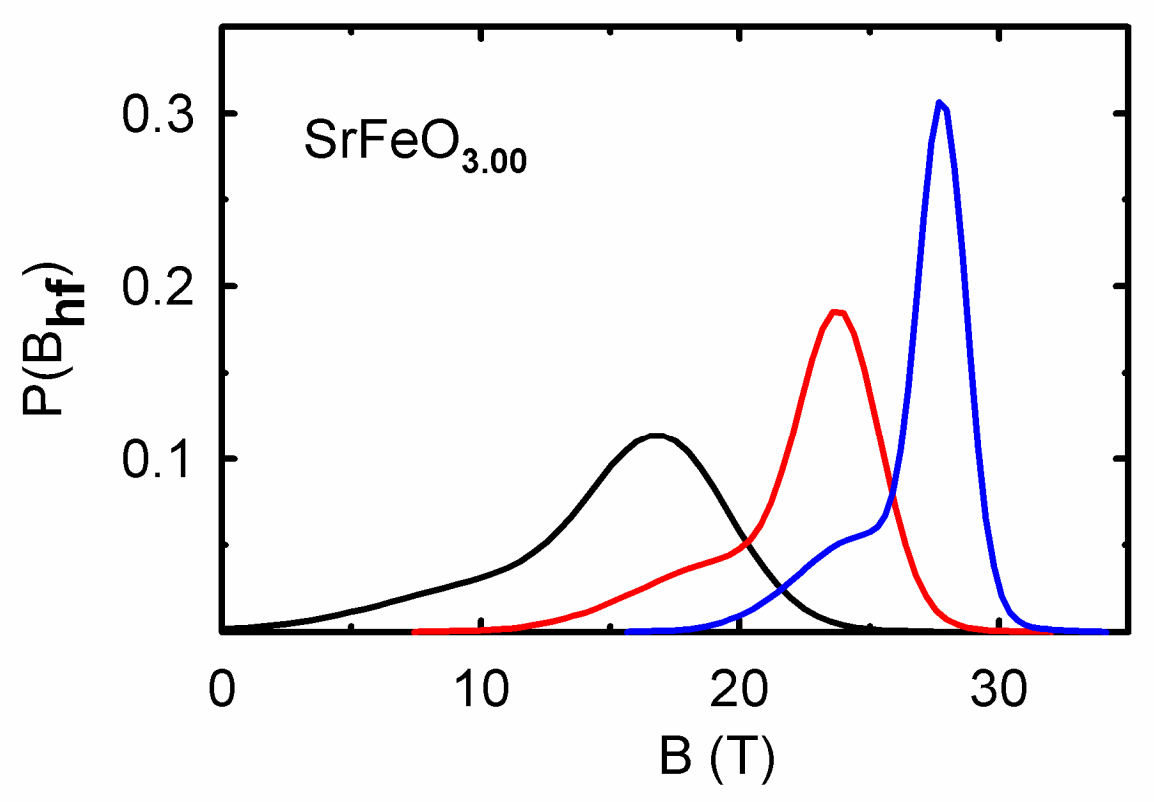

FIG. 4

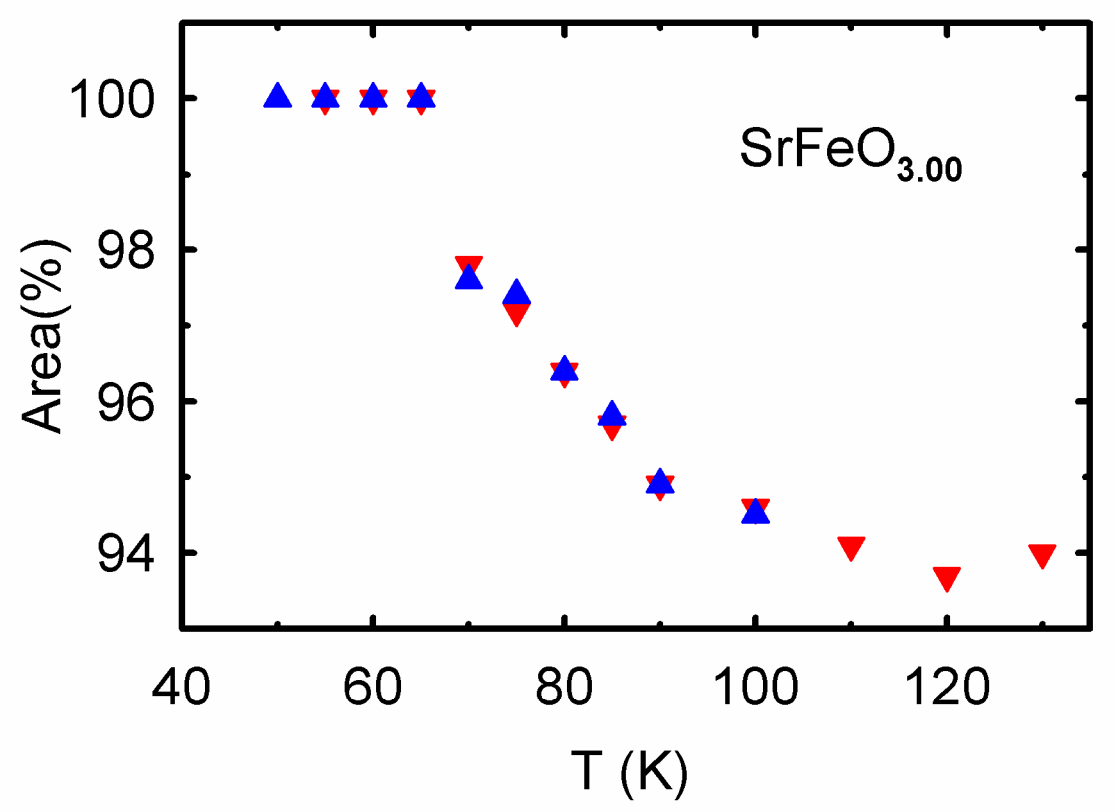

FIG. 5

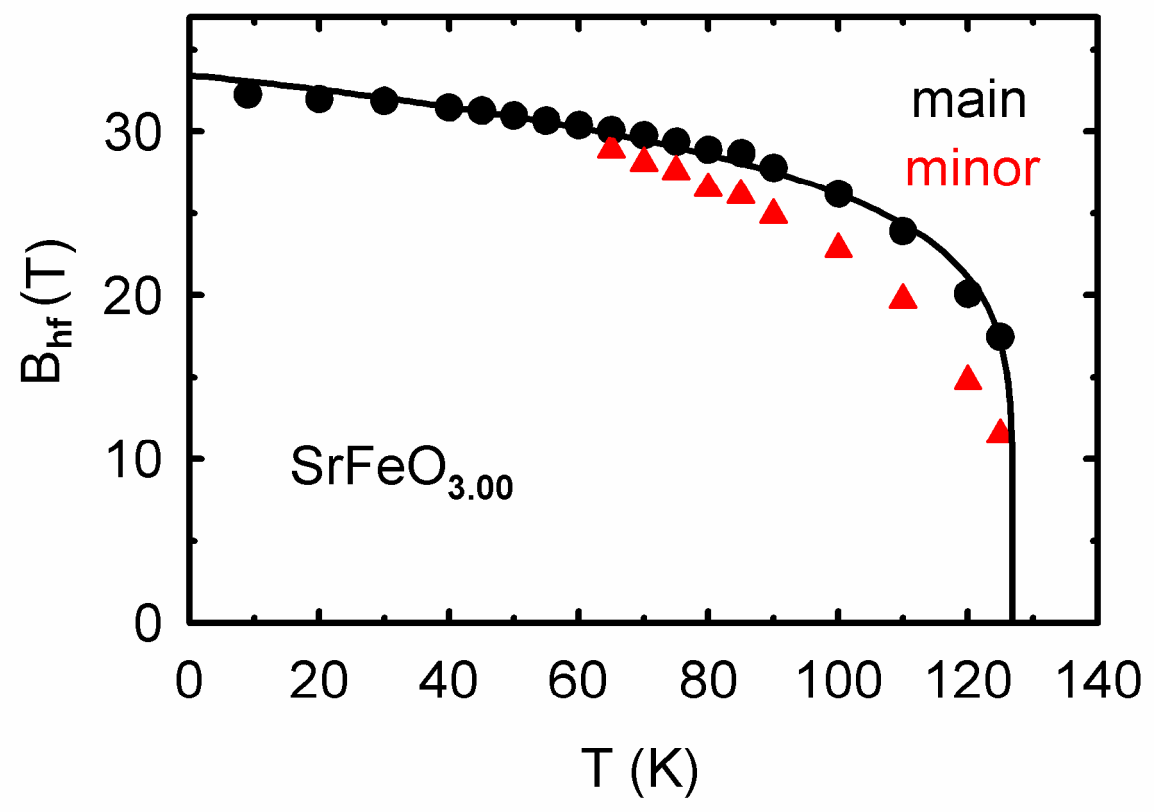

FIG. 6

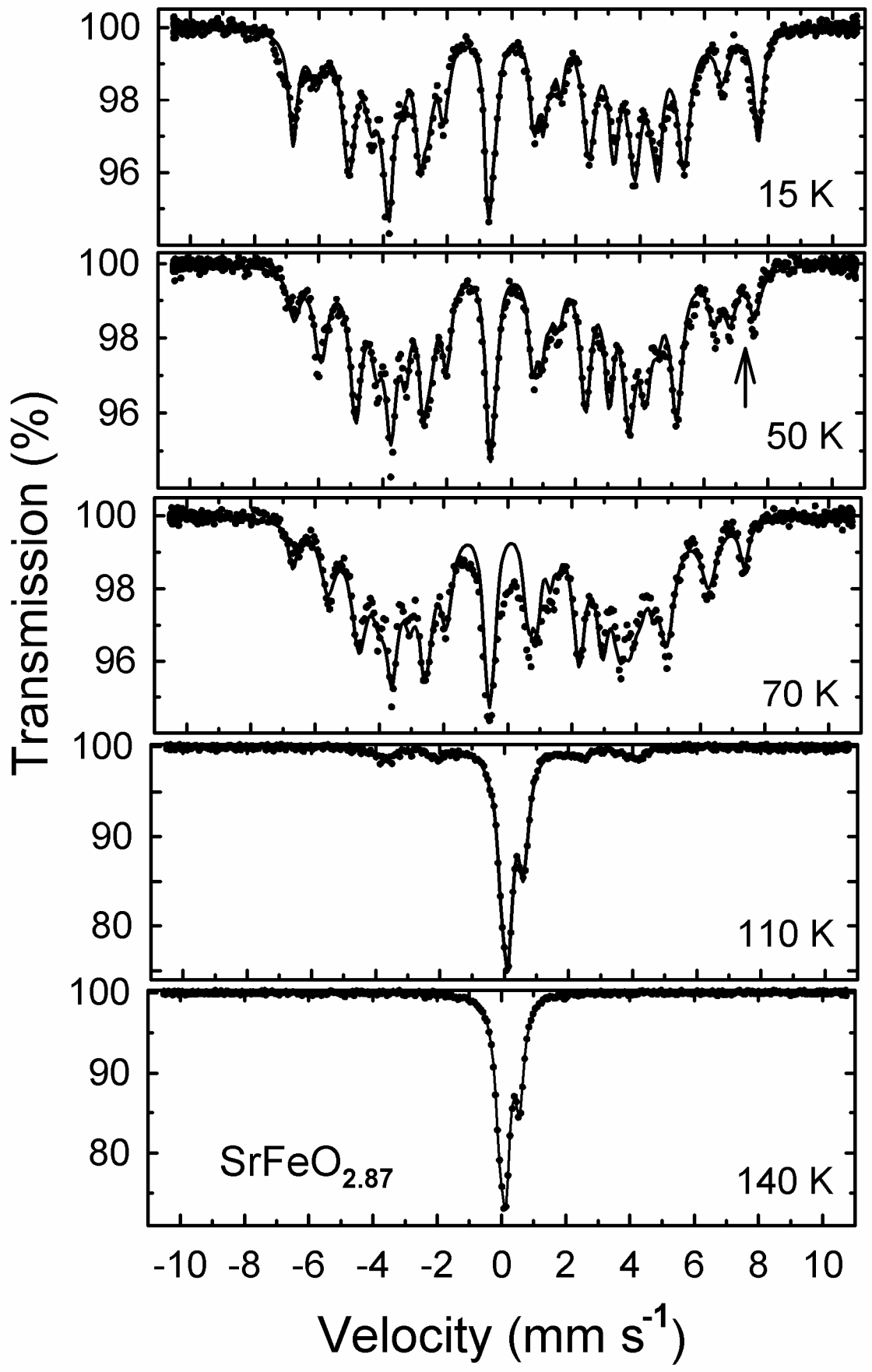

Transmission (%)
100
98
96
15 K
100
98
96
50 K
100
98
96
70 K
100
90
80
110 K
100
90
80
SrFeO2.87
140 K
-10 -8 -6 -4 -2 0 2 4 6 8 10
Velocity (mm s-1)

FIG. 7

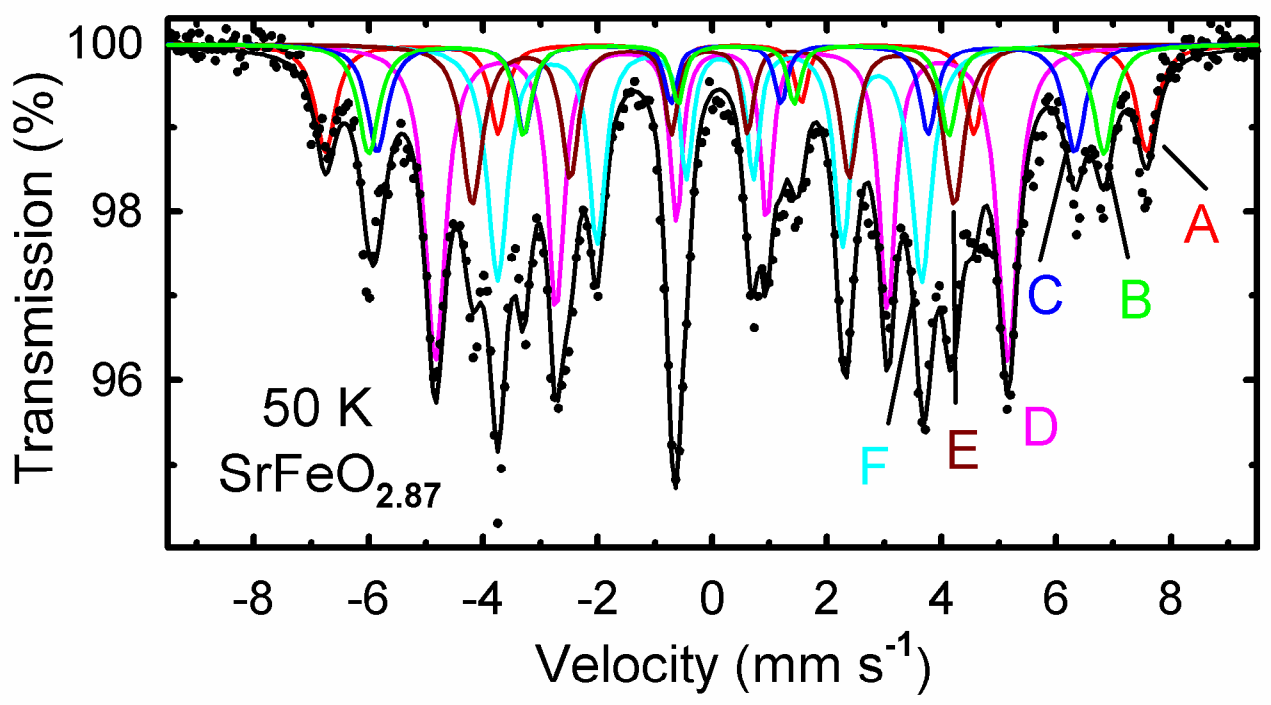

FIG. 8

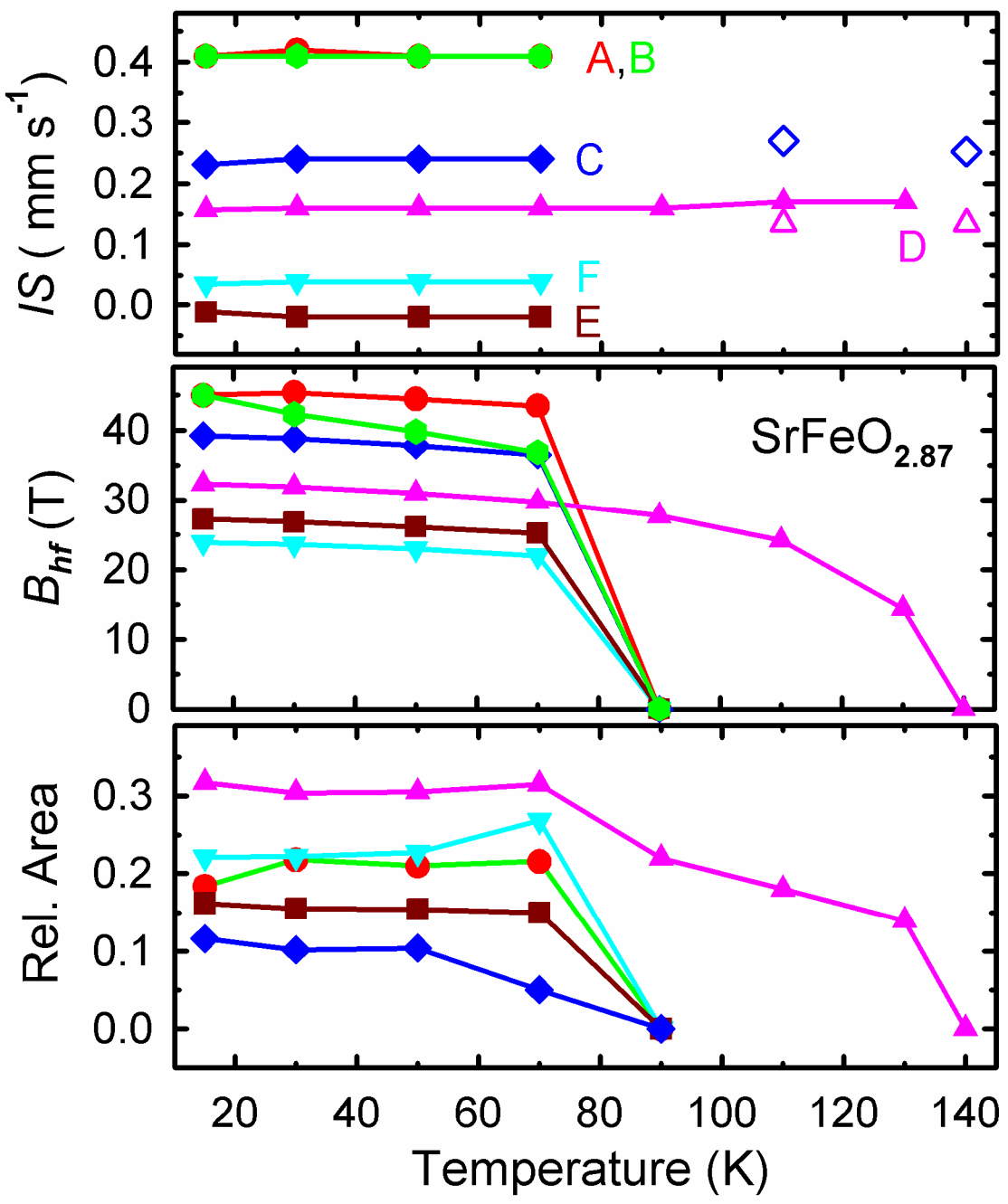

FIG. 9

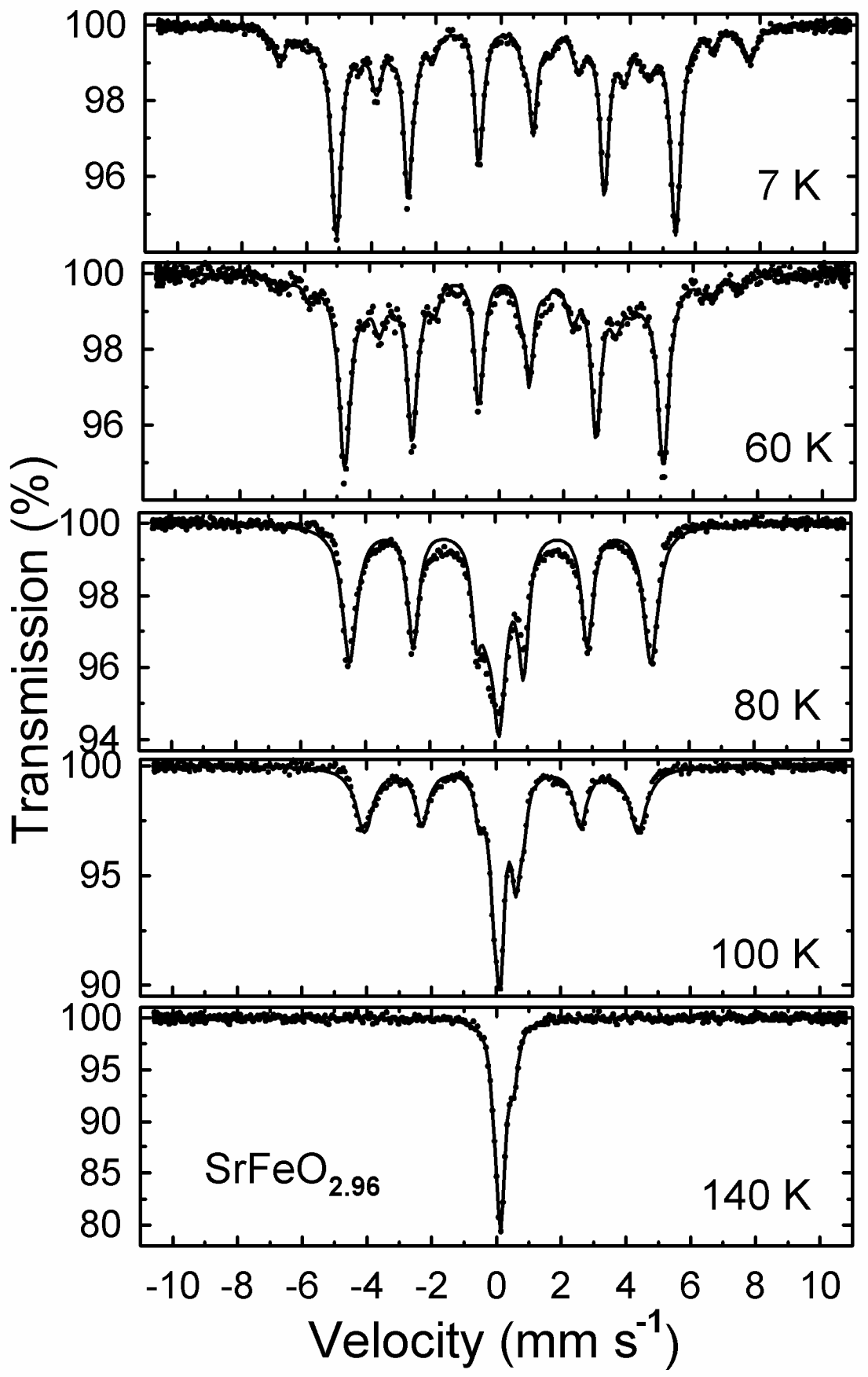

FIG. 10

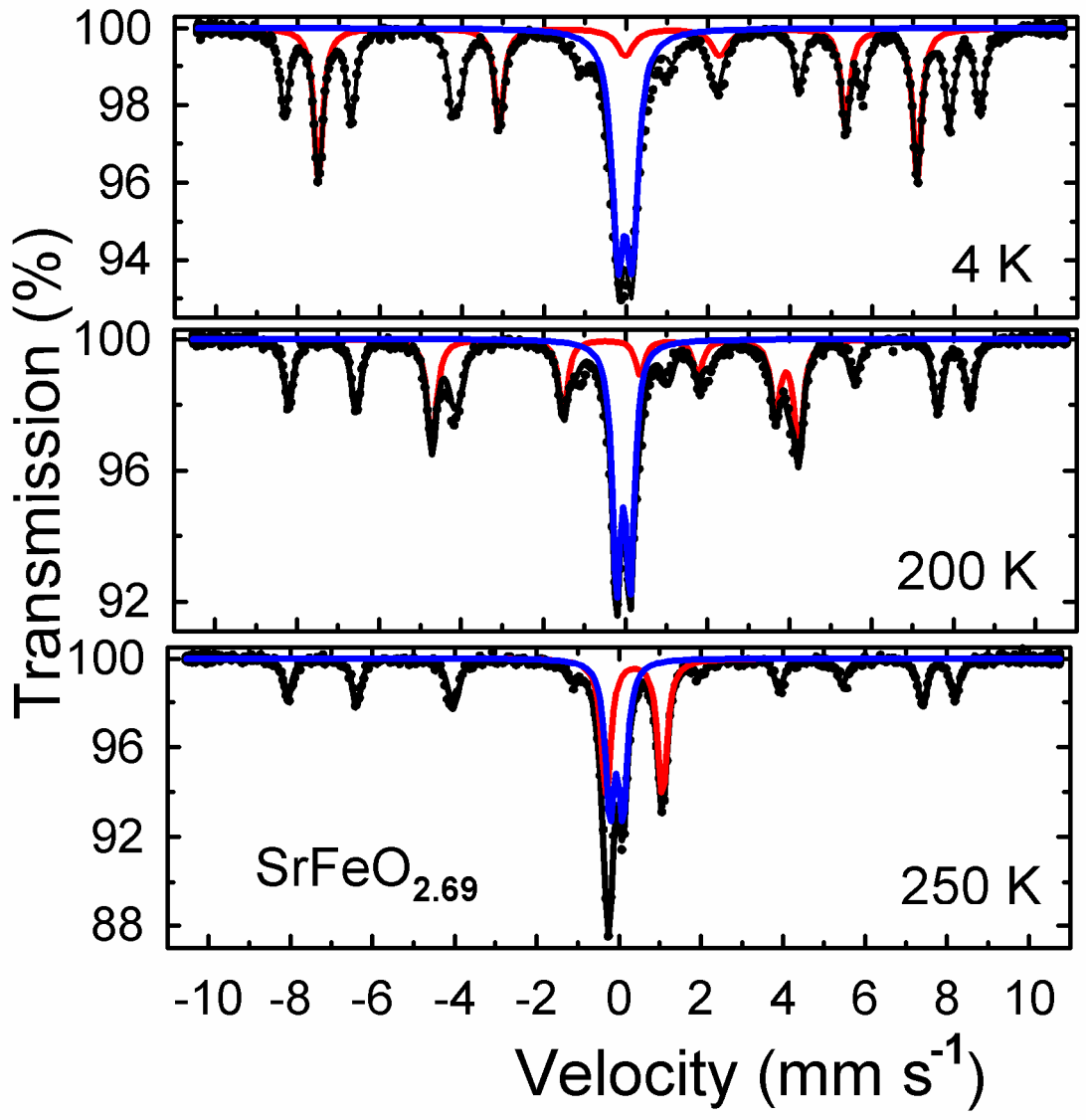

100
98
96
94
4 K
100
96
92
200 K
100
96
92
88
SrFeO2.69
250 K
Transmission (%)
-10 -8 -6 -4 -2 0 2 4 6 8 10
Velocity (mm s-1)

FIG. 11

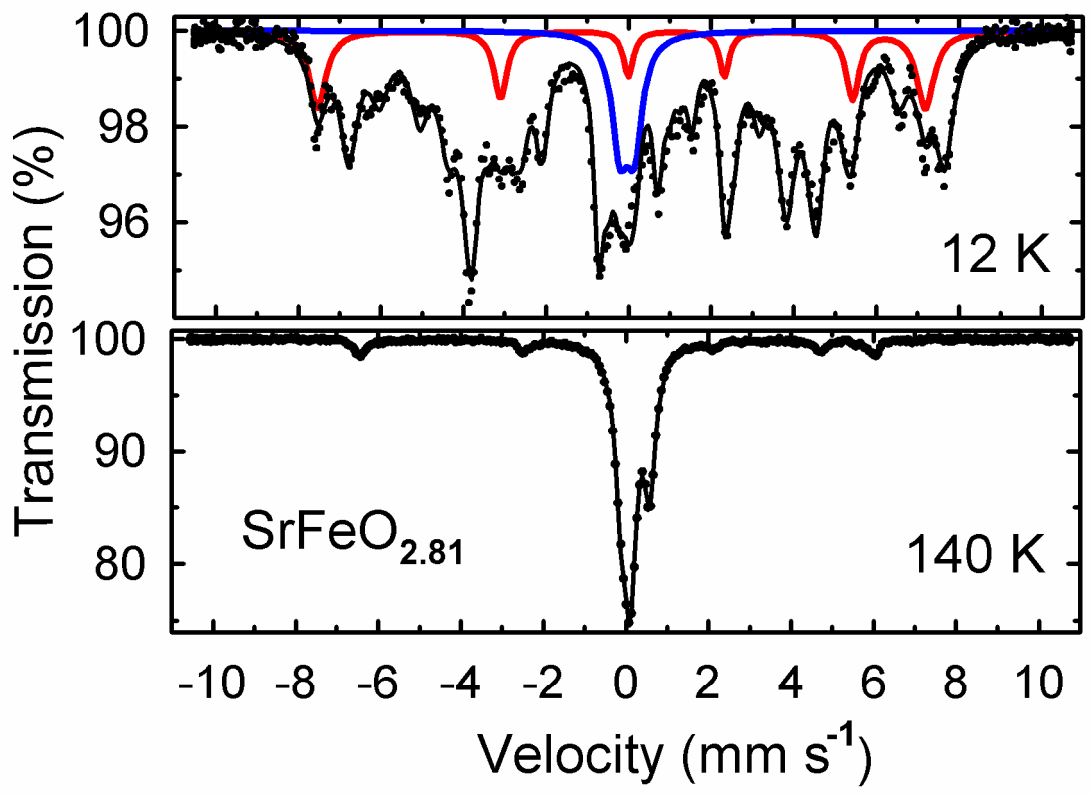

100
98
96
12 K
100
90
80
Transmission (%)
SrFeO2.81
140 K
-10 -8 -6 -4 -2 0 2 4 6 8 10
Velocity (mm s-1)

FIG. 12

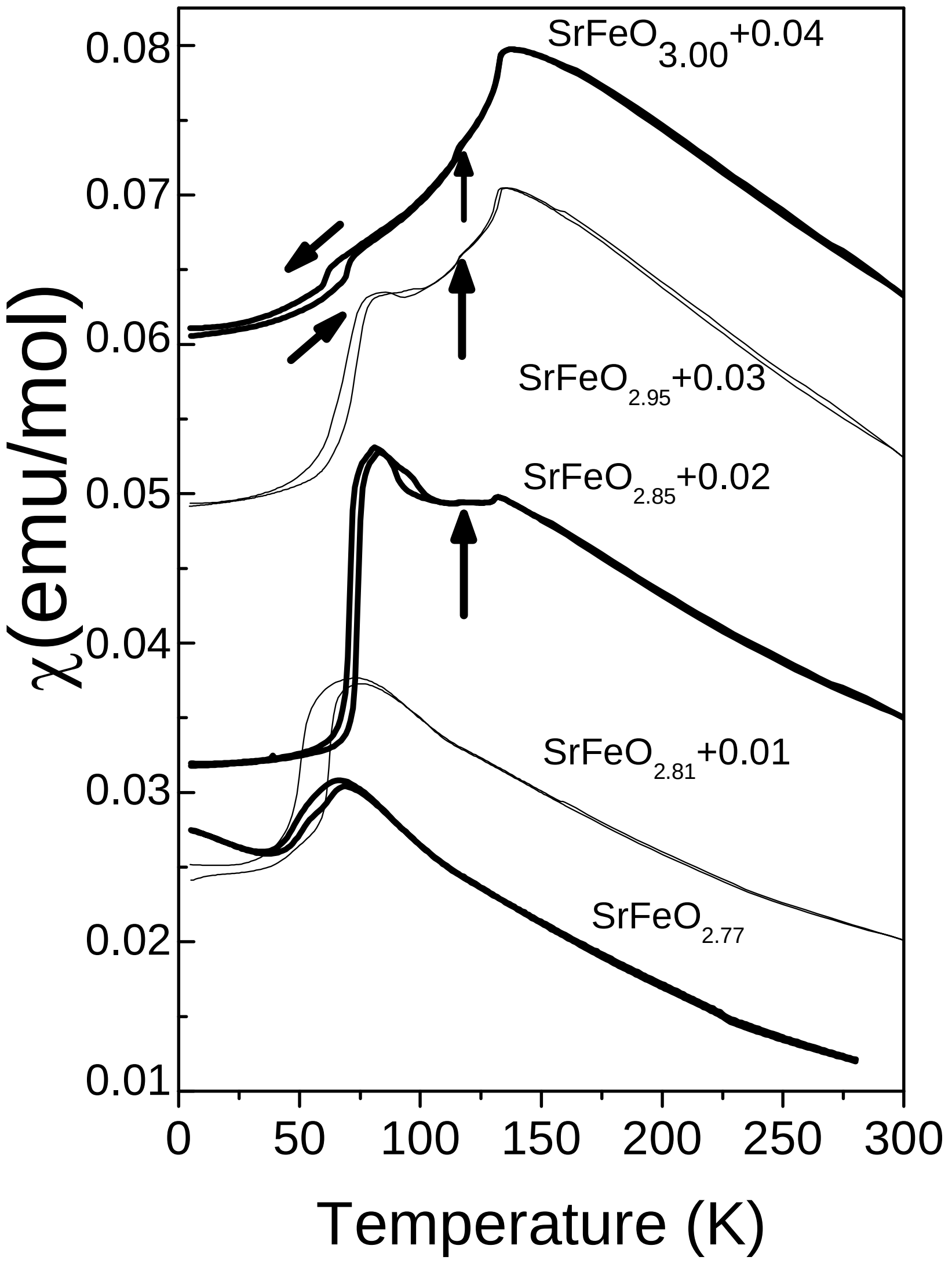

FIG. 13

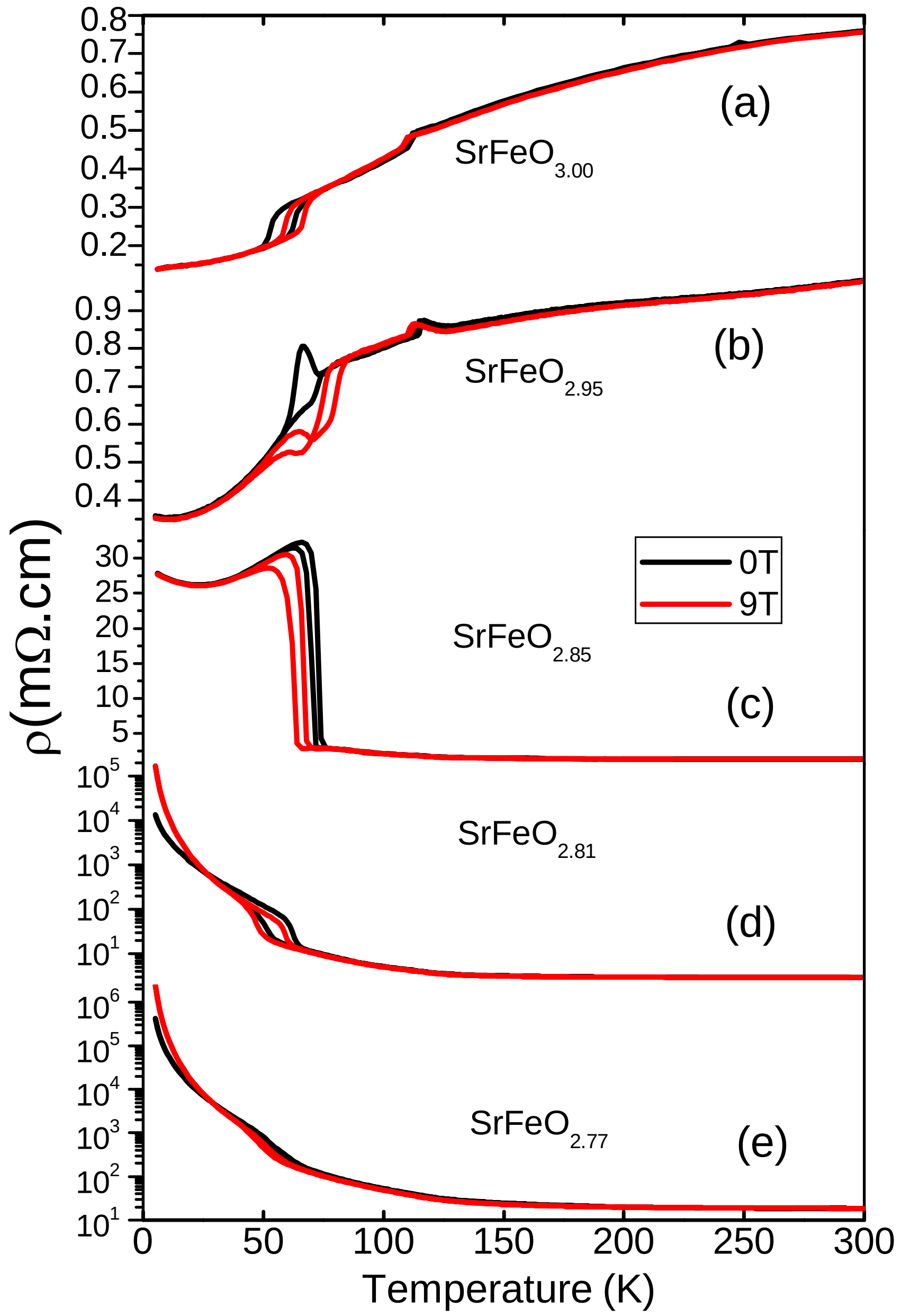

FIG. 14

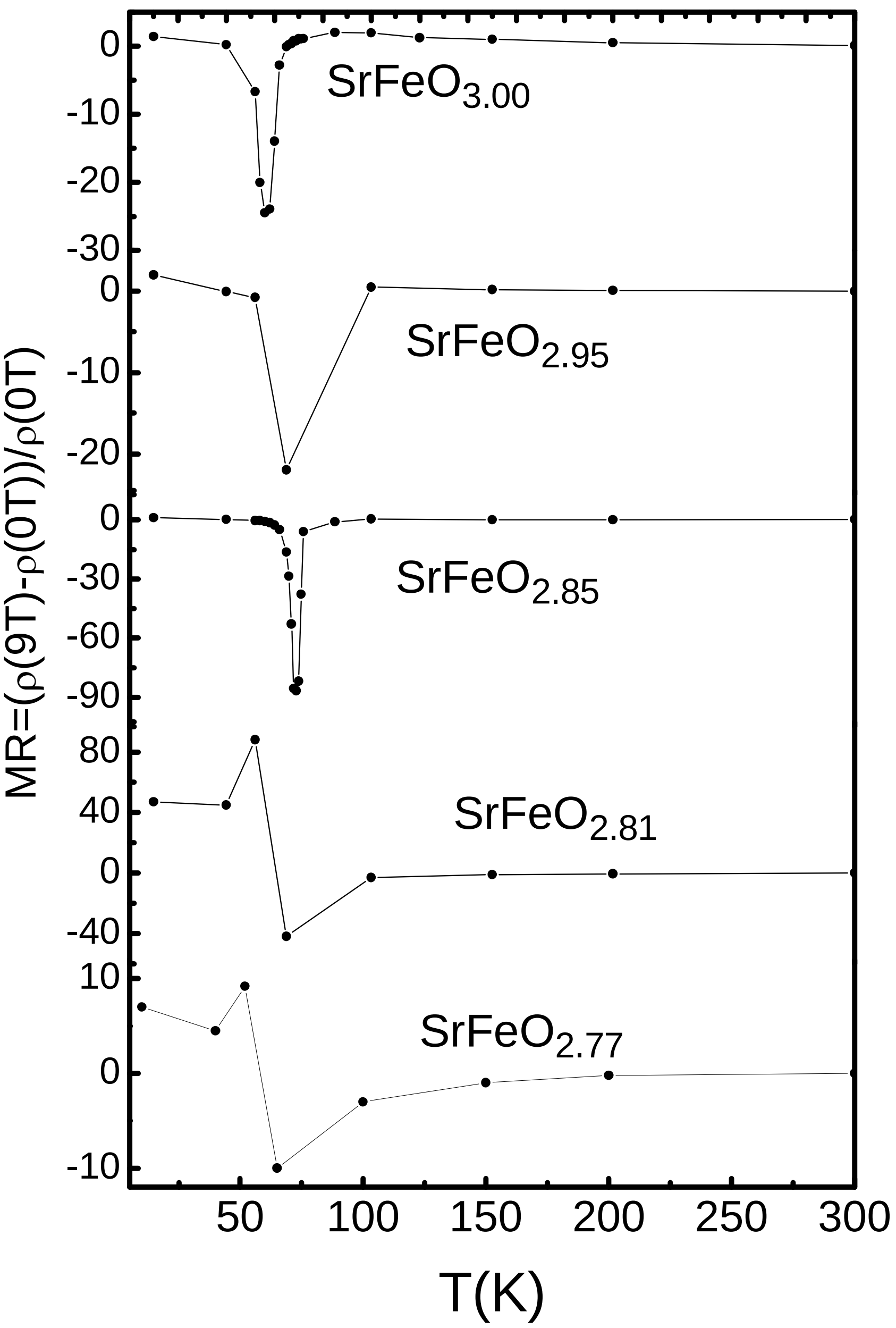

FIG. 15

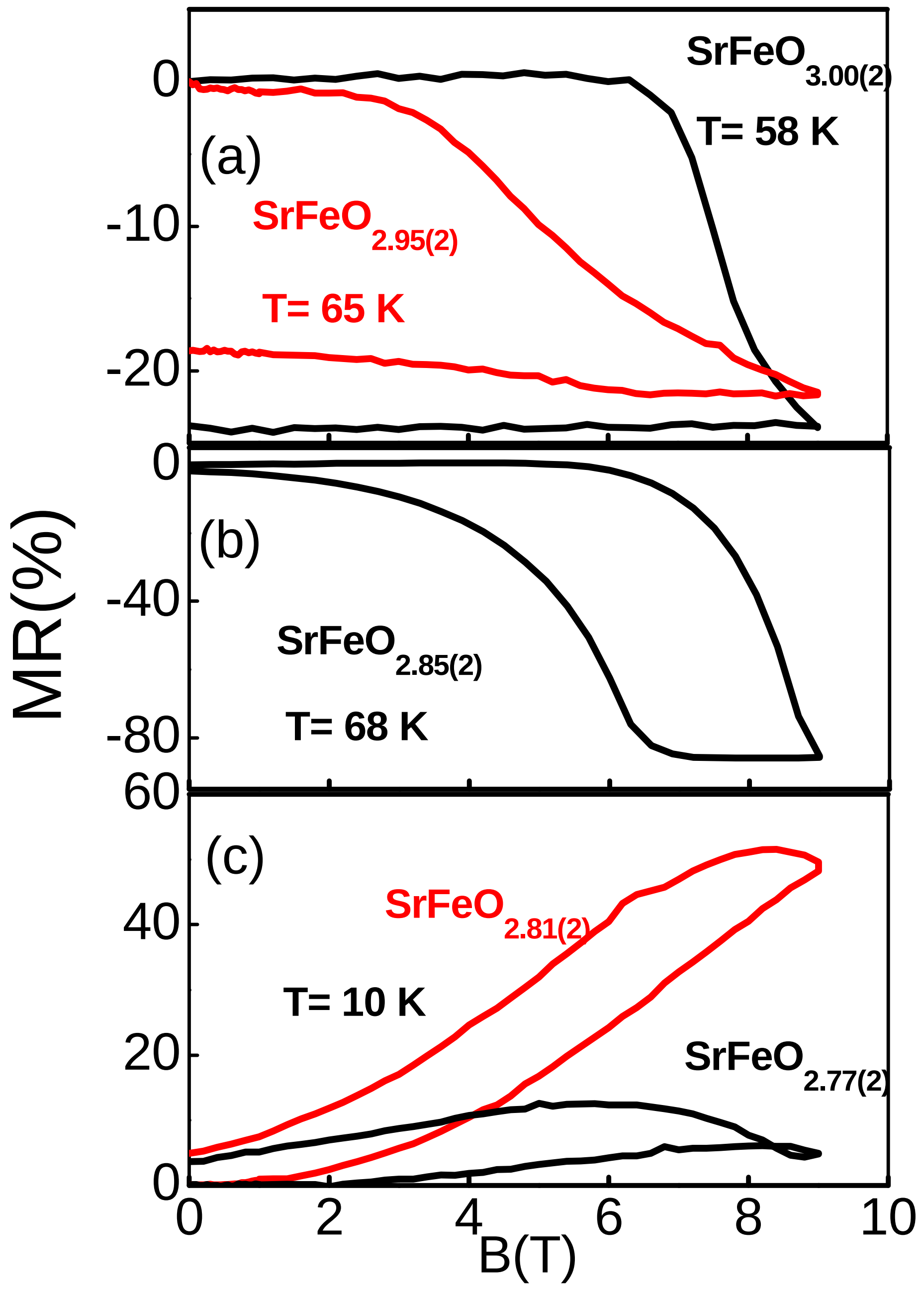

FIG. 16

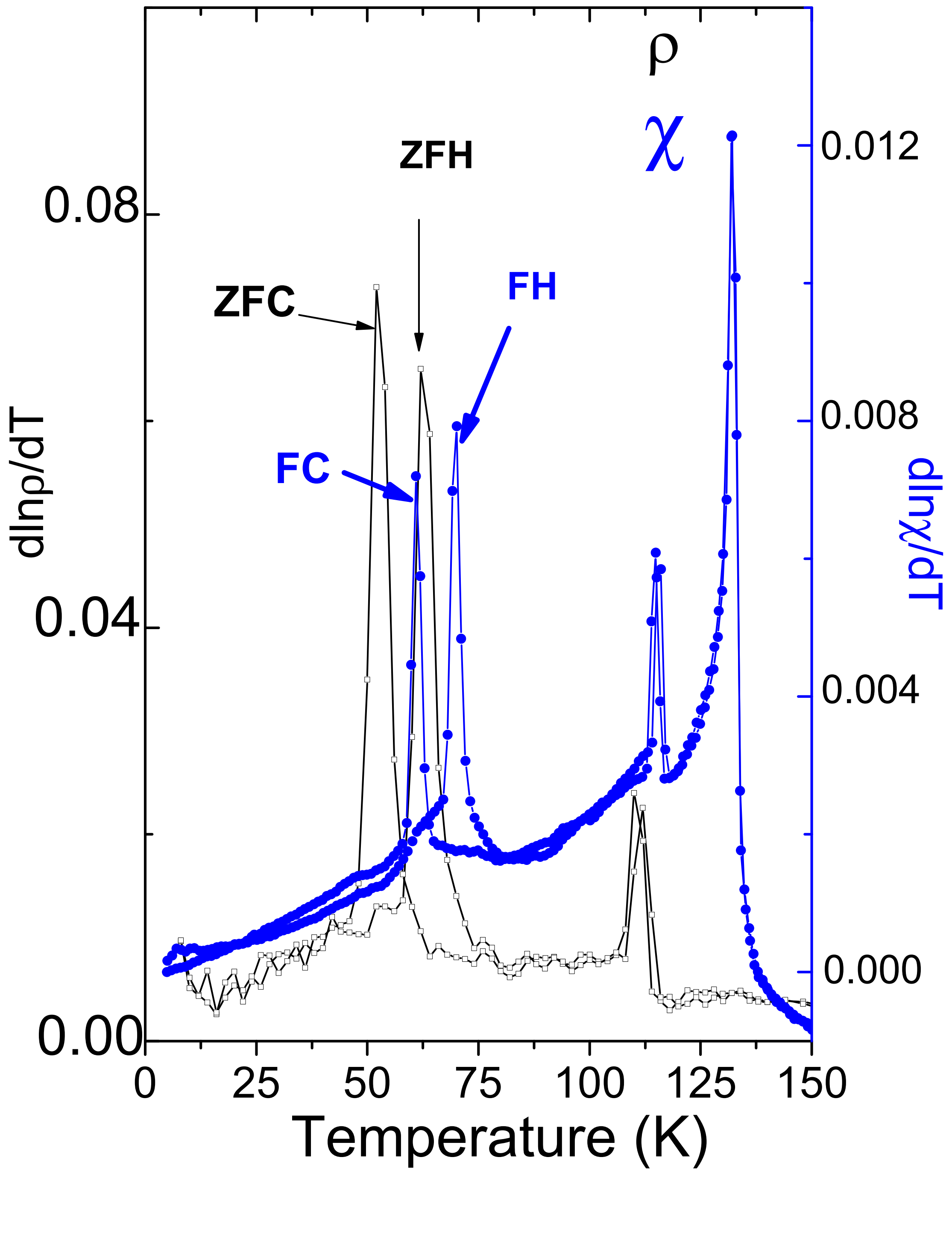

Fig. 17

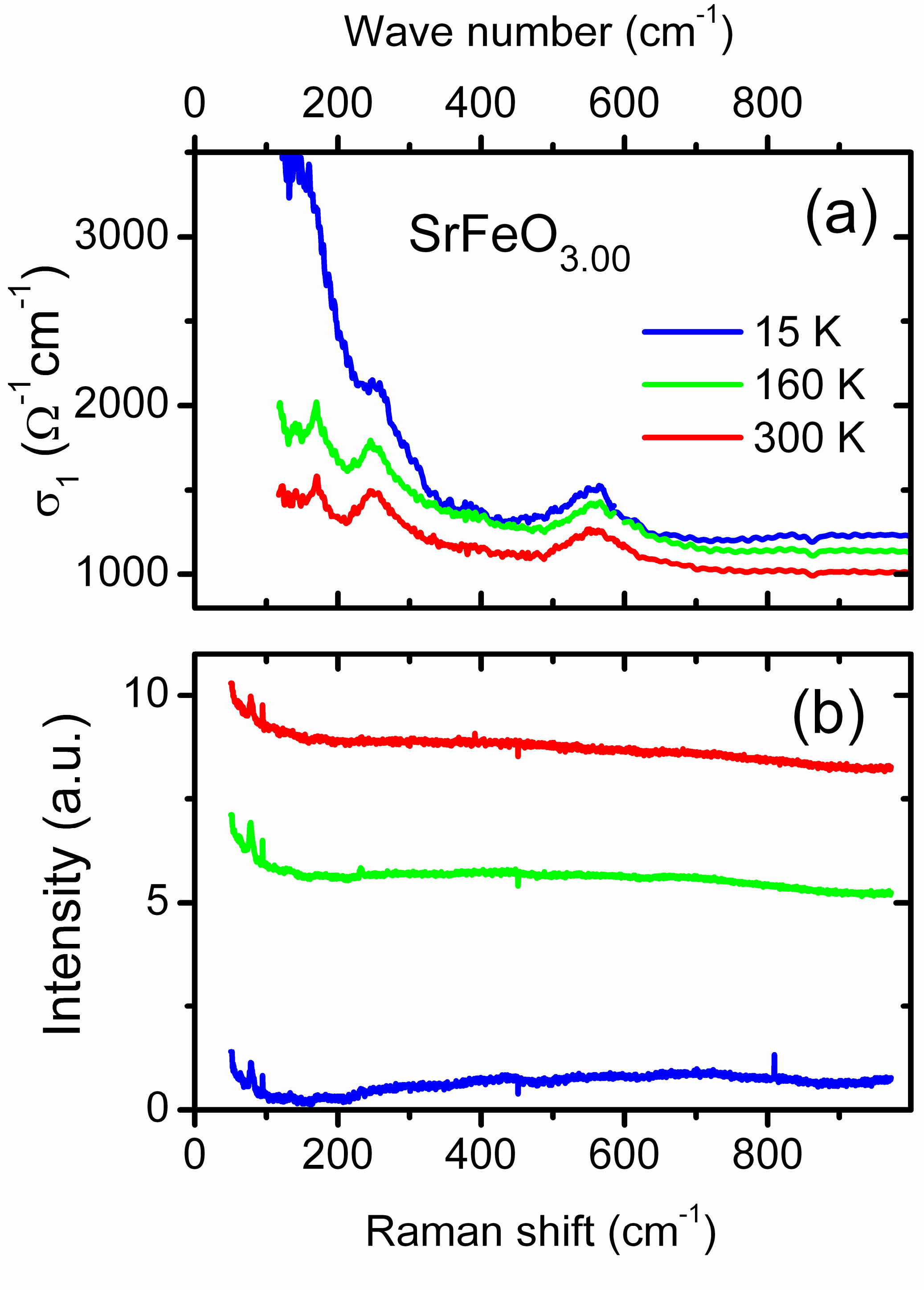

Fig. 18

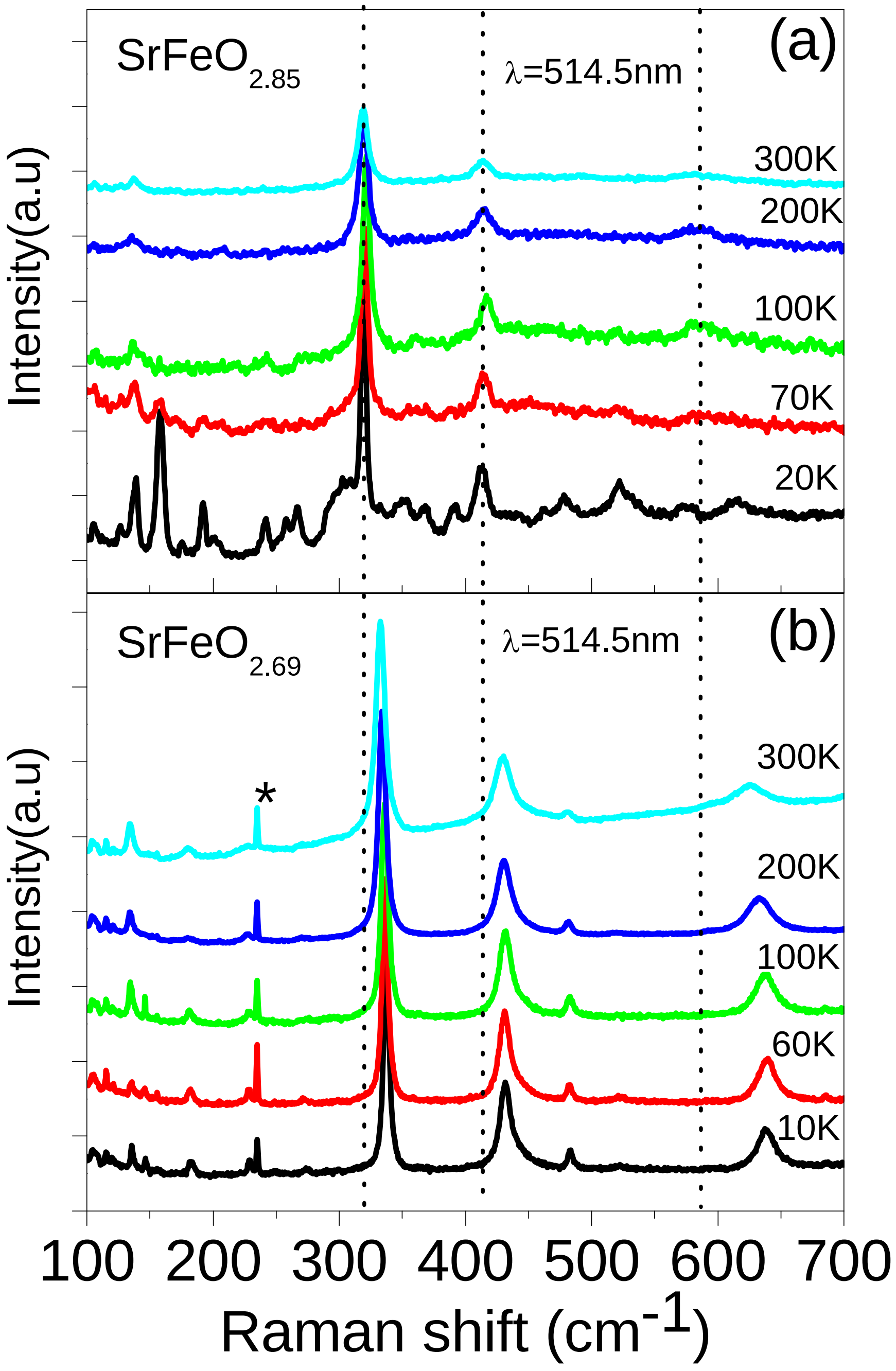

FIG. 19

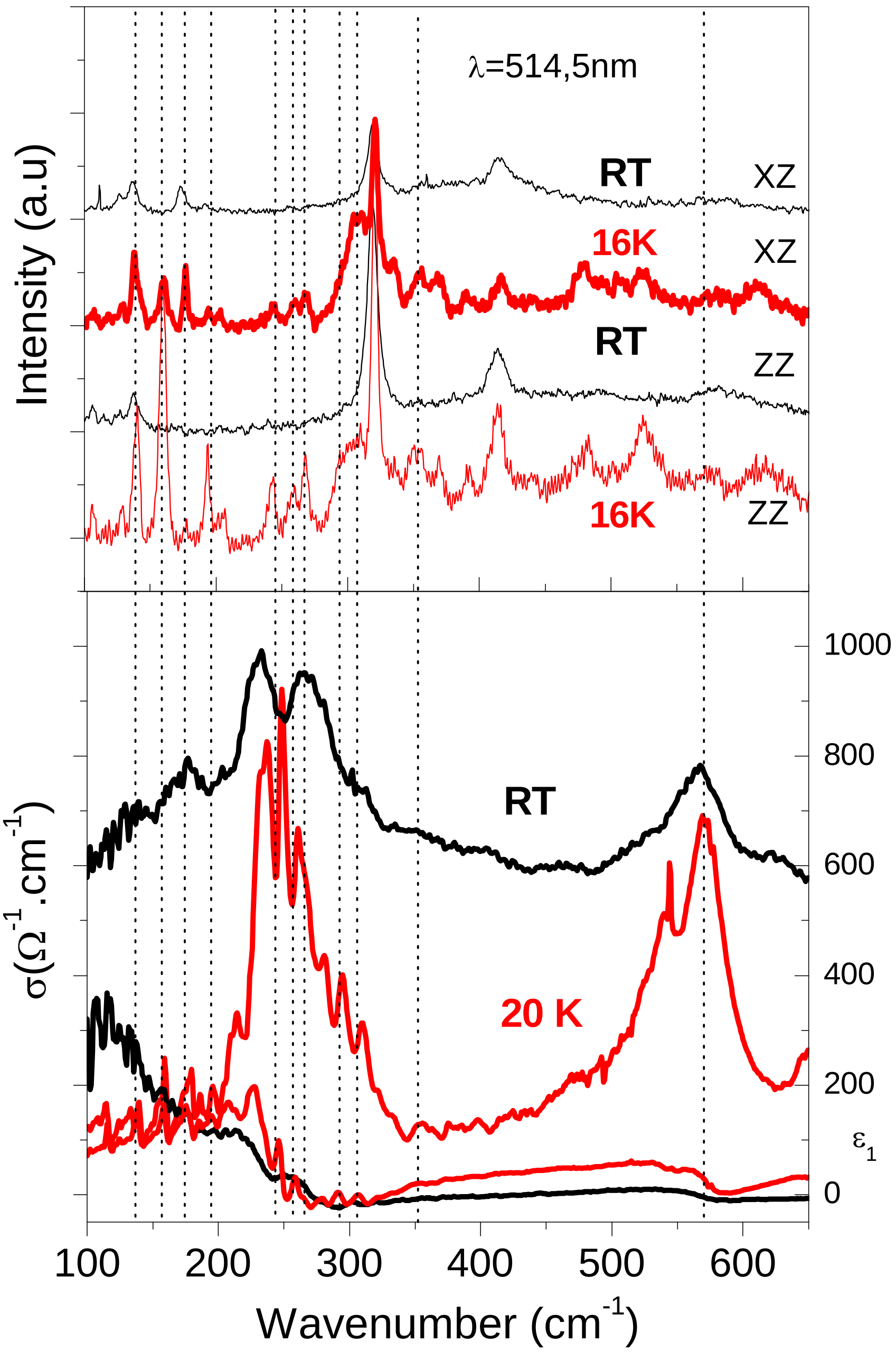

FIG. 20